\documentclass[aps,twocolumn,preprintnumbers,prb,amsmath,
amssymb,amsfonts,superscriptaddress,floatfix]{revtex4}

\usepackage{amsfonts}
\usepackage{amssymb}
\usepackage{amsmath}
\usepackage{ulem}
\usepackage{color}
\usepackage{latexsym}
\usepackage{graphicx}
\usepackage{subfigure}
\usepackage{graphics}
\usepackage{floatflt}
\usepackage{epsfig}
\usepackage{overpic}
\usepackage{soul} 
\usepackage{varioref,xr-hyper}
\usepackage[latin1]{inputenc}
\usepackage{mathbbol}
\usepackage{bbm}

\newcommand{\red}{\textcolor{black}}

\newcommand{\cyan}{\textcolor{black}}

\newcommand{\be}{\begin{equation}}
\newcommand{\ee}{\end{equation}}
\newcommand{\bea}{\begin{eqnarray}}
\newcommand{\eea}{\end{eqnarray}}

\def\nn{\nonumber}
\def\lb{\label}
\def\pref#1{(\ref{#1})}

\def\ra{\rightarrow}

\def\Ra{\Rightarrow}

\def\bk{{\bf k}}

\def\bq{{\bf q}}

\def\bQ{{\bf Q}}

\def\bS{{\bf S}}

\def\a{\alpha}
\def\b{\beta}

\def\d{\delta}

\def\t{\tau}

\def\s{\sigma}
\def\o{\omega}
\def\G{\Gamma}
\def\D{\Delta}
\def\O{\Omega}

\newcount\bozza \bozza=0
\ifnum\bozza=1
\newdimen\shift \shift=-1.0truecm
\def\lb#1{%
{\label{#1}\rlap{\kern\shift{$\scriptstyle#1$}}}}
\else\def\lb#1{\label{#1}} \fi

\begin{document}
\title{Anisotropy of the dc conductivity due to orbital-selective spin 
fluctuations in the nematic phase of iron superconductors}

\author{Raquel Fern\'andez-Mart\'{i}n}
\affiliation{Materials Science Factory, Instituto de Ciencia de Materiales de Madrid, 
ICMM-CSIC, Cantoblanco, E-28049 Madrid, Spain}
\author{Laura Fanfarillo}
\email{laura.fanfarillo@sissa.it}
\affiliation{CNR-IOM and International School for Advanced Studies (SISSA), Via
Bonomea 265, I-34136, Trieste, Italy}
\author{Lara Benfatto}
\affiliation{ISC-CNR and Dep. of Physics, ``Sapienza'' University of Rome, P.le
A. Moro 5, 00185, Rome, Italy} 
\author{Bel\'en Valenzuela}
\email{belenv@icmm.csic.es}
\affiliation{Materials Science Factory, Instituto de Ciencia de Materiales de Madrid, 
ICMM-CSIC, Cantoblanco, E-28049 Madrid, Spain}

\date{\today}
\begin{abstract} {We study the DC conductivity of iron-based superconductors 
within the orbital-selective spin fluctuation scenario. Within this approach, 
the anisotropy of spin fluctuations below the spin-nematic transition at $T_s$ 
is also responsible for the orbital ordering, induced by nematic self-energy 
corrections to the quasiparticle dispersion. As a consequence, the anisotropy of 
the DC conductivity below $T_s$ is determined not only by the anisotropy of 
the scattering rates as expected within a spin-nematic scenario, but also by the 
modification of the Fermi velocity due to the orbital reconstruction. More 
interestingly,  it turns out that these two effects contribute to the 
DC-conductivity anisotropy with opposite signs. By using realistic 
band-structure parameters we compute the conductivity anisotropy for both 122 
and FeSe compounds, discussing the possible origin of the different 
dc-conductivity anisotropy observed experimentally in these two families of 
iron-based superconductors. }\end{abstract}

{\bf \maketitle }

The driving force of electronic nematicity is one of the most intriguing puzzles 
in iron-based superconductors (IBS). The structural transition from tetragonal 
to orthorhombic at $T_S$ comprised the nematic phase characterized by a marked 
electronic anisotropy, much larger than the one expected by the structural 
transition itself\cite{GallaisReview16}. In most iron pnictides the structural 
transition precedes or coincides with the magnetic transition at $T_N$.  The 
proximity of the magnetic phase led to the proposal of the band spin-nematic 
scenario\cite{schmalianprb12,fernandesnatphys14} where the spin fluctuations 
(SF) at $\bold{Q}_X=(\pi,0)$ and at $\bold{Q}_Y=(0,\pi)$ become anisotropic 
below $T_S$. The lack of long-range magnetic order in FeSe has cast some doubts 
on the validity of the spin-nematic scenario in this compound. FeSe presents a 
nematic phase below the structural transition at $T_S=90K$ down to the critical 
superconducting temperature $T_c\sim 9K$. Even though a magnetic phase is not 
stabilized in temperature, sizeable SF have been detected also in FeSe 
\cite{Rahn15, Wang2016, Meingastprb18, Furukawaprb17, FurukawaarXiv18}. ARPES 
experiments in the nematic phase report a momentum-modulated orbital splitting 
between the $\Gamma$ and M point of the Brillouin zone 
\cite{ColdeaWatsonreview18} that has been interpreted via both the 
orbital-ordering scenario \cite{BaekNatMat14, OnariKontaniprl16, SuJcondMat15, 
MukherjeePRL15, Jiangprb16, Chubukovprb17} and the spin-nematic scenario 
\cite{Fanfarilloprb16}. In this situation two related questions arise: what is 
the role of the spin-orbital interplay and whether the origin of nematicity is 
universal in IBS or material-dependent\cite{Chubukovprx16, Kontaniprx16, 
Fanfarillo2018}. 

Resistivity anisotropy is a hallmark of nematicity in IBS. In detwinned 
electron(e)-doped 122 compounds $\Delta \rho = \rho_x-\rho_y<0$ is found below 
the structural transition\cite{mazin10, degiorgi10, fisher2011, degiorgi2012, 
mirriprb14} while detwinned hole(h)-doped compounds present the opposite 
anisotropy\cite{prozorovnatcomm13}. There is an on-going debate in the 
literature on whether the observed DC anisotropy is due to the anisotropy in the 
scattering rate or to the anisotropy in Fermi Surface (FS) 
parameters\cite{sciencedavis10, nakajimaprl12, Uchidaprl13, davisnatphys2013, 
fisherprl14, FisherDegiorgiprl15, Degiorgiprb16, FernandesDCACprb16, 
Chinottiprb17}. In principle, within an orbital-ordering scenario the different 
occupation of the various orbitals affects mainly the  
FS\cite{leeyinku09,lvphillipsprb10,kontaniprb11}, while within a spin-driven 
scenario the largest effect is expected to come from an anisotropy in the 
inelastic scattering rate\cite{Fernandesprl10, sciencedavis10, timmprb14, 
andersenprl14, Hirschfeldprl15} . Specifically, in the band spin-nematic 
scenario, depending on the FS shape and size, the band nesting is active at the 
so called hot spots on the FS, where the scattering rate is maximum. It has been 
argued that the location of the hot spots could explain the different signs 
between e-doped compounds and h-doped compounds\cite{prozorovnatcomm13} in 
pnictides. Besides the spin-nematic or orbital order scenario, further attempts 
to explain the DC anisotropy in pnictides taking into account the spin-orbital 
interplay has been performed using effective spin-fermion model\cite{liangprl12} 
or multiorbital microscopic model in the magnetic phase\cite{nosotrasprl10-2, 
Bascones16, sugimotoprb14}.

Recent experiments in FeSe have found the opposite anisotropy with respect to 
the e-doped 122 compounds\cite{Prozorovprl16}, i.e. $\Delta \rho = 
\rho_x-\rho_y>0$. Given the significant FS reconstruction observed in the 
nematic phase of FeSe\cite{ColdeaWatsonreview18}, we need to revise the role of 
the scattering rate and velocity anisotropies taking into account the 
spin-orbital interplay in order to theoretically address both pnictides and 
FeSe. Within an orbital-ordering scenario, the opposite anisotropy of the 
resistivity of 122 and FeSe compounds in the nematic phase has been ascribed to 
the orbital-dependent inelastic quasiparticle scattering by orbital-dependent 
SF\cite{Kontaniprb17}. However, an analogous study of the DC conductivity 
anisotropy within a spin-nematic scenario accounting for the spin-orbital 
interplay and able to address pnictides and FeSe, is still missing.

\begin{figure}[thb] 
\includegraphics[width=0.64\linewidth,clip=true]{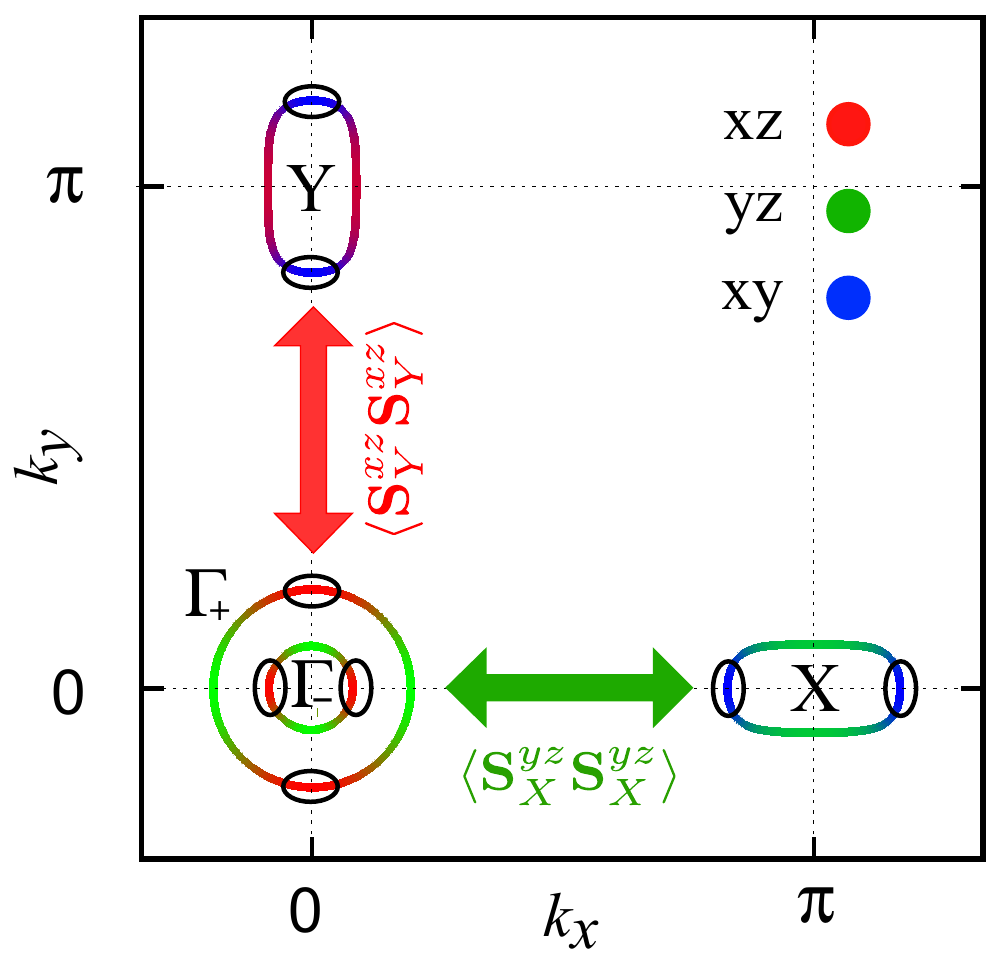} 
\caption{General
sketch of the orbital content of the Fermi surface of the 4-pocket model for
iron superconductors. The green and red arrows show the OSSF with $yz$ content
in the x-direction and $xz$ content in the y-direction. Cold spots, where
the scattering rate is minimum, are shown by a circle and they are found on the
$xy$ and $xz$ orbitals in the nematic phase due to anisotropic self-energy
corrections. See text.} 
\vspace{-0.35cm}
\label{fig-OSSF} 
\end{figure}

The aim of this work is then to provide an interpretation for the observed 
differences, using as a starting point the orbital-selective spin fluctuation 
(OSSF) model. The model, derived in the itinerant approach\cite{Fanfarillo2018}, 
exploits the original idea of the orbital selective character of the SF in IBS 
discussed in Ref.~[\onlinecite{Fanfarilloprb15}]. Within the OSSF model, due to 
the orbital composition of the FS the SF peaked at $\bold{Q}_X=(\pi,0)$ involve 
only the $yz$ orbital, while the SF at $\bold{Q}_Y=(0,\pi)$ involve only the 
$xz$ orbital, Fig.~\ref{fig-OSSF}. As in the spin-nematic scenario, the nematic 
phase emerges when SF at $\bold{Q}_X=(\pi,0)$ and at $\bold{Q}_Y=(0,\pi)$ become 
anisotropic, however in the OSSF model such anisotropy directly affects the 
$yz/xz$ orbital symmetry. The OSSF is a minimal model that explains successfully 
the enhanced nematic tendency of FeSe as compared to 122 
systems\cite{Fanfarillo2018} and clarifies controversial experimental issues in 
FeSe such as the temperature evolution of the FS of FeSe and the odd orbital 
ordering observed by ARPES experiments\cite{Fanfarilloprb16}, the decrease of 
the nematic critical temperature and the emergence of magnetism in FeSe with 
pressure\cite{Fanfarillo2018} as recently observed in\cite{Sunnatcomm17, 
Kothapallinatcomm17, ColdeaWatsonreview18}. Moreover the analysis of the 
superconductivity mediated by anisotropic OSSF\cite{FanfarilloarXiv18} 
successfully account for the enigmatic anisotropy of the superconducting gap 
revealed by STM\cite{Davis2017} and ARPES\cite{BorisenkoarXiv18, Rhodesprb18, 
Zhouarxiv18} experiments in FeSe. 

In this work we analyze, within the OSSF model, the effect of anisotropic
self-energy corrections on the conductivity anisotropy in the nematic phase of
IBS. In contrast to the band spin-nematic scenario\cite{schmalianprb12,
fernandesnatphys14}, where just the scattering rate contributes to the
conductivity anisotropy, we found that also the velocity contributes. The
contribution of the scattering rate to the resistivity anisotropy is dominated
by the location of the cold spots where the scattering rate is minimum (see
Fig.~\ref{fig-OSSF}), which, within our model, is determined by the orbital
composition of the FS and by the spin-orbital interplay of the OSSF. The
contribution of the velocity to the resistivity anisotropy is counter-intuitive
and opposite to the one of the scattering rate. We find indeed that the
conductivity is larger in the direction where the self-energy is also larger.
This interesting new effect is due to an orbital character exchange in the
pockets arising from the OSSF self-energy in the nematic phase. Our study shows
that the sign of the anisotropy of the DC conductivity depends on whether
scattering rate or velocity anisotropy dominates on each pocket, as well as
other parameters such as the ellipticity and the quasiparticle renormalization
due to local interactions\cite{demedicichapter}. Thus, different experimental
results among the various families of IBS can be explained within the same OSSF
scenario. 

The structure of the paper is the following. In sec.\,I we introduce the OSSF
model. In sec.\,II we outline the calculation of the DC conductivity: in
sec.\,II\,A we derive analytical expressions for the DC anisotropy within the
perfectly-nested parabolic-band approximation; in sec.\,II\,B we discuss
numerical results obtained using realistic parameters for 122 compounds and FeSe
i.e. accounting for the effects of spin-orbit coupling and elliptical e-pockets;
in sec.\,II\,C we discuss our results in connection to experiments in IBS. In
sec.\,III we summarize our results and draw the conclusions of our work.

\section{Model}

The OSSF low--energy model has been derived in detailed in 
Ref.\,[\onlinecite{Fanfarillo2018}]. Here we summarize the main features of the 
model, further details can be found in App.\,A. 
The starting point is a general 4-pocket model with two h-pockets at $\G$, 
denoted as $\G^\pm$, and two e-pockets at $X$ and $Y$. The model 
can be easily adapted to describe different compounds among the 122 and 11 
families. The kinetic part of the Hamiltonian is derived following the 
low-energy approach considered in Ref.\,[\onlinecite{Vafekprb13}], where each 
pocket is described using a spinor representation in the orbital space. 
\be 
\lb{h0}
H_0^l=\sum_{\bk,\s} 
\psi^{\dagger l}_{\bk\s} \hat H_{0 \bk}^l \psi^l_{\bk\s},\;\;
\hat H^l_{0 \bk}= h_{0 \bk}^l \hat{\t_0} + \vec{h}^l_{\bk} \cdot \vec{\hat{\t}}
\ee
with $l=\G,X,Y$ and $\hat{\t}$ the Pauli matrices representing the orbital 
pseudo-spin. The spinors are defined as $\psi^{\Gamma}_{\bk 
\s}=(c^{yz}_{\bk,\s},c^{xz}_{\bk \s})$ and $\psi^{X/Y}_{\bk \s}=(c^{yz/xz}_{\bk 
\s},c^{xy}_{\bk \s})$. Rotating the Hamiltonian into the band basis we have

\be
\lb{h0_band}
H_0^l=\sum_{\bk,\s} E^{l \pm}_{\bk} c^{\dagger l\pm}_{\bk \s} c^{l\pm}_{\bk \s},
\ee
where $E^{l \pm}_{\bk} = h_{0 \bk}^l\pm h^l_{\bk}, \quad 
h^l_{\bk}=|\vec{h}_{\bk}^l|$ are the band dispersions.  The fermionic band 
operators $c^{l\pm}$ are obtained rotating the orbital spinors  via an 
unitary matrix ${\hat{\cal U}}^{l}$. Explicitly for the h-pockets at $\Gamma$, 
$c^{\G\pm} \equiv h^{\pm}$ we have 
\be
\lb{eq:Umatrix}
\begin{pmatrix}
h^+ \\
h^- \\
\end{pmatrix} = 
\begin{pmatrix}
u^{\Gamma} &-v^{\Gamma}\\
v^{*\Gamma} & u^{*\Gamma}\\
\end{pmatrix}
\begin{pmatrix}
 c^{yz} \\
 c^{xz} \\
\end{pmatrix} ={\hat{\cal U}}^\G
\begin{pmatrix}
 c^{yz} \\
 c^{xz} \\
\end{pmatrix}
\ee
where we have dropped the momentum and spin indices for simplicity. Analogous 
expressions hold for the $X/Y$ e-pockets fermionic operators $c^{X/Y\pm} 
\equiv e^{X/Y \pm}$. Since only the $E^{X/Y +}$ band crosses the Fermi level at 
$X/Y$, in the following we will drop the $+$ subscript from $e^{X/Y}$. 

The interacting Hamiltonian simplifies substantially once the spin-exchange 
interaction is projected at low-energy \cite{Fanfarilloprb15, Fanfarillo2018}. 
The generic intraorbital spin operator reads $\bS^\eta_{\bq }=\sum_{\bk \a 
\b}(c^{\eta \dagger}_{\bk \a} \vec \s _{\a \b}c^\eta_{\bk+\bq \b})$, with $\eta$ 
the orbital index and $\vec{\s}_{\a \b}$ the Pauli matrices for the spin. Thus, 
taking into account the orbital composition of the FS shown in 
Fig.~\ref{fig-OSSF}, the relevant intraorbital spin operator occur at momenta 
$\bq$ near ${\bf Q}_X=(\pi,0)$ and ${\bf Q}_Y=(0,\pi)$, which connect a h-pocket 
with the $X$ and $Y$ e-pockets respectively. Since the only common orbital is 
the $yz/xz$ along the $x/y$ direction the spin interaction reduces to
\be
\lb{hint_low}
H_{int}= - \frac{\tilde{U}}{2} \sum_{\bq \\'} \bS^{yz/xz}_{X/Y} \cdot \bS^{yz/xz}_{X/Y}.
\ee
Here, $\tilde{U}$ is the intraorbital interaction renormalized at low energy and 
$\bS_{X/Y}^\eta\equiv \vec S_{\bq=\bQ_{X/Y}}^\eta$. The relevant magnetic 
fluctuations peaked at $\bQ_X$/$\bQ_Y$ are orbital selective, as sketched in 
Fig.\ \ref{fig-OSSF}, having only $yz$/$xz$ orbital character:
\bea
\lb{corrx}
\langle \bS\cdot \bS\rangle (\bQ_X) &\Ra& \langle \bS^{yz}_{X}\cdot \bS^{yz}_{X}\rangle \\
\lb{corry}
 \langle \bS\cdot \bS\rangle (\bQ_Y) &\Ra& \langle \bS^{xz}_{Y}\cdot \bS_{Y}^{xz}\rangle   
\eea

The SF exchange between h- and e-like pockets renormalizes the quasiparticles
via single-particle self-energy corrections. Within the OSSF model, due to the
orbital-selective nature of SF, this mechanism is also orbital dependent. The
self-energy corrections in the orbital basis can be computed within an
Eliashberg-like treatment\cite{Fanfarilloprb16} for each pockets. 
For the h-pockets at $\G$ we find 
\be
\lb{smatrixG}
\hat \Sigma^\G (\o)=\begin{pmatrix}
\Sigma^\G_{yz}(\o) & 0\\
0 &\Sigma^\G_{xz}(\o) \\
\end{pmatrix}=\Sigma^\G_0(\o)\hat\t_0+\Sigma_3^\G(\o)\hat\t_3,
\ee
with $\Sigma^\G_0(\o)=(\Sigma^\G_{yz}(\o)+\Sigma^\G_{xz}(\o))/2 $, 
$\Sigma_3^\G(\o)=(\Sigma^\G_{yz}(\o)-\Sigma^\G_{xz}(\o))/2 $, 
while for the e-pockets we have
\be
\lb{smatrixXY}
\hat \Sigma^{X/Y}(\o)=\begin{pmatrix}
\Sigma^{X/Y}_{yz/xz} (\o)& 0\\
0 &0 \\
\end{pmatrix}=\Sigma^{X/Y}_0(\o)\hat{\t_0}+\Sigma_3^{X/Y}(\o)\hat{\t_3},
\ee
with $\Sigma^{X/Y}_0(\o)=\Sigma_3^{X/Y}(\o)=\Sigma^{X/Y}_{yz/xz}(\o)/2$. In the 
tetragonal state, above the structural transition, the 
isotropic SF lead to equivalent self-energies for the $C_4$ symmetric 
$xz/yz$ orbitals, i.e. $\Sigma^\G_3= 0$, and $\Sigma^{X}_3=\Sigma^{Y}_3$. In the 
nematic phase the anisotropy of the OSSF below $T_S$ generates a differentiation 
of the $xz/yz$ self-energy corrections
 \bea
 \lb{eq:SelfepocketNem}
 \Sigma^\G_3(\o)&\ne &0 \nn\\
 \Sigma^{X}_3(\o)&\ne&\Sigma^{Y}_3(\o),
\eea
and gives rise to an effective orbital nematicity.
Eq.s~\pref{smatrixG}-\pref{smatrixXY} are generic but their specific values are
material dependent and for FeSe were calculated at the RPA level in
Ref.~[\onlinecite{Fanfarilloprb16}]. 
The dressed Green's functions are obtained via the Dyson equation $\hat
G^{l^{-1}}(\bk,\o)=\hat G_0^{l^{-1}}(\bk,\o)-\hat \Sigma^l(\o)$ where $\hat
G_{0}^{l^{-1}}(\bk,\o)=\o\hat{\mathbb{1}}-\hat{H}_0^l(\bk)$. We diagonalize the
renormalized Green's function via the unitary transformation ${\hat{\cal
U}}_{R}^l(\bk,\o)$ defined as 
\be
\lb{grot}
\hat{G^l}(\bk,\o)={\hat{\cal U}_R^l}(\bk,\o)(\o\hat{\mathbb{1}}-
\hat{\Lambda}_{R}^l(\bk,\o))^{-1}{\hat{\cal U}}_R^{l^{-1}}(\bk,\o). 
\ee
where $\hat{\Lambda}_{R}^l = diag( E^{l+}_R, E^{l-}_R)$ and
\bea
\lb{ER}
 E_{R}^{l_\pm}(\bk,\o)&=& h_{\bk 0}^l+\Sigma_0^l(\o)\pm h^l_{R\bk,\o} \nn \\ 
  h^l_{R\bk,\o}&=&\sqrt{(h_{3}^l+\Sigma_3^l(\o))^2+(h^l_{1})^2+(h^l_{2})^2}. 
\eea
Since the self-energy is a complex function, Eq.\ \pref{ER} accounts both for 
the renormalization of the band dispersion 
\be
\lb{ReR}
\epsilon_{R}^{l_\pm}(\bk,\o)  = Re E_{R}^{l_\pm}(\bk,\o)
\ee
and for the renormalized scattering rate 
\be
\lb{GR}
\G_R^{l_\pm}(\bk,\o) = \d \G +|Im E_{R}^{l_\pm}(\bk,\o)|
\ee
where we also added a residual constant broadening term $\d \G$. 
Eq.~\pref{GR} establishes a connection between the scattering rate and the
self-energy renormalizations contained in $E_{R}^{l_\pm}$. The qualitative
behavior of the self-energies in the nematic phase allows us to easily localize
the minimum value of $\G_R^{l_\pm}$ on the FS, i.e. the cold spots shown in
Fig.~\ref{fig-OSSF}. As we discussed in Ref.\ [\onlinecite{Fanfarilloprb16}],
the reconstruction of the FS below $T_S$ is consistent with the Ising-nematic
spin-fluctuations being bigger at $\bold{Q}_X$ than at $\bold{Q}_Y$. This
implies that self-energy corrections are stronger on the $yz$ orbital than on
the $xz$ one. As a consequence on the h-pockets the smaller scattering rate
corresponds to the $xz$ orbital. On the e-pockets instead, the smaller
scattering rate is found for the $xy$ orbital, given the absence of $xy$-SF
within our model. The result is an example of the spin-orbital interplay
retained by the OSSF approach that allows us to directly link the cold spots
position with the FS orbital character and is not present in the 
band-based spin-nematic scenario\cite{schmalianprb12}.  

The rotation matrix ${\hat{\cal U}_R^l}(\bk,\o)$ in Eq.~\pref{grot}, has a
structure analogous to Eq.~\pref{eq:Umatrix}. The renormalized coherence
factors, $u^l_{R}$ and $v^l_{R}$, whose exact expressions are given in App.\,A, 
depend now on the self-energies Eq.s \pref{smatrixG}-\pref{smatrixXY}. The
approximated expressions for $u^l_{R\bk}(\o)$, $v^l_{R\bk }(\o)$ obtained at low
energy $\o$ by expanding in $\Sigma^l_3$ at the first order read:
\bea
\lb{uvrenexpand}
\begin{split}
|u^l_{R}(\o)|^2=&|u^l|^2 \left(1 + \frac{2{Re}\Sigma_3^l(\o)}{h^l}|v^l|^2\right), \\
|v^l_{R}(\o)|^2=&|v^l|^2 \left(1 - \frac{2{Re}\Sigma_3^l(\o)}{h^l}|u^l|^2\right)
\end{split}
\eea
where we neglected the imaginary part of the self-energy which approaches to 
zero at low energy and low temperature. $u^l$ and $v^l$ are the bare 
coherence factors appearing in Eq.\pref{eq:Umatrix} and detailed in App.\,A. 
From Eq.~\pref{uvrenexpand} one sees that the correction term $\sim 
{Re}\Sigma_3^l(\o)$ mixes the orbital character in each pocket, i.e. contribute 
to $u^l_{R}$ with a term proportional to $v^l$ and viceversa. This effect of the 
OSSF self-energy in the coherence factors will have important consequences for 
the renormalized velocities as we will see in the following section.

\section{DC conductivity}

The DC conductivity is the $\Omega \to 0$ limit of the longitudinal optical
conductivity given by 
\be
\lb{eq:cond}
\text{Re}\,\sigma_{\a}(\O)=-\frac{e^2}{V}\frac{\text{Im}\, 
\Pi_{\a}(\bq=0, i\O_m \ra \O)}{\O},
\ee
where $\alpha=x,y$, $V$ is the unit-cell volume and $\Pi_\a$ is the 
current-current correlation function that in the bubble approximation reads
\bea
\Pi_\a(\bq, {i\O_m})&=& 2T \sum_{l \bk n} \text{Tr}
\Big[ \hat{G^l}(\bk -\bq/2,i\o_n) \hat {V}_{\bk_\a}^l  \nn \\
&& \hat G^l (\bk+\bq/2 ,i\o_n+i\O_m) \hat {V}_{\bk_\a}^l  \Big] .
\lb{pi}
\eea
$\hat{V}_{\bk_\a}^l=\partial_{\bk_\a} \hat H_0^l$ is the bare velocity operator
and $\hat{G^l}(\bk,i\o_n)$ is the renormalized Green's function defined in
Eq.~\pref{grot}.  We rotate Eq.\pref{pi} in the band basis using
Eq.~\pref{grot}, perform the trace and take the $\Omega \ra 0$ limit. The
$\a=x,y$ component of the DC conductivity in the band basis, 
${\sigma_{DC}}_{\a} = \sum_{l_\pm} \sigma^{l_\pm}_\a$, is obtained from the
sum over all the pockets $l_\pm=\G_+, \G_-, X,Y$. The pocket conductivity 
reads (see App.\,B for further details)
\be
\lb{eq:DC}
\sigma^{l_\pm}_{\a} = \frac{2 \pi e^2}{ N}\sum_{\bk}\int_{-\infty}^{\infty}
d\omega(-\frac{\partial f(\omega)}{\partial\omega}) 
\big({V}_{R\bk_\a }^{l_\pm}(\o)\big)^2 \big( A^{l_\pm}_\bk(\o)\big)^2.
\ee
In this basis the spectral function is diagonal 
\be
\lb{spectralf}
A^{l_\pm}_\bk(\o)=\frac{1}{\pi}\frac{\G_R^{l_\pm}(\o)}{(\G_R^{l_\pm}(\o))^2 + 
(\omega - \epsilon_{R\bk}^{l_\pm}(\o))^2}
\ee
with $\G_R^{l_\pm}(\o)$ and $\epsilon_{R\bk}^{l_\pm}(\o)$ defined in 
Eq.s~\pref{ReR}-\pref{GR}. $V^{l_\pm}_{R \bk\a }$ in Eq.s~\pref{eq:DC} is the bare 
velocity operator rotated into the band basis. As a consequence of the orbital 
structure we have 
\be
\lb{vel}
V^{l_\pm}_{R \bk \a}= V_{\bk\a}^{l_{11}} |u^l_{R}|^2 
\pm V_{\bk\a}^{l_{12}} u^{*l}_{R}v^{*l}_{R} \pm V_{\bk\a}^{l_{21}}
u^l_{R}v^l_{R} +V_{\bk\a}^{l_{22}} |v^l_{R}|^2
\ee
Hereafter we omit the dependence on $\o$ for simplicity. ${V_{\bk \a}^{l_{\eta 
\eta^{\prime}}}}$ are the $\eta \eta^{\prime}$ component of the velocity 
and $({u/v})^l_{R}$ are the renormalized coherence factors. Via the coherence 
factors $V^{l_\pm}_{R \bk \a}$ depends on the $\hat{\tau}_{3}$ component of 
the self-energies, $\Sigma_3^l(\o)$ that mixes the orbital content of each 
pocket. This effect can be easily understood considering the approximated 
expressions in Eq.s~\pref{uvrenexpand}. By neglecting the imaginary 
part of the self-energy in $({u/v})^l_{R}$, Eq.~\pref{vel} can also be written as 
\be
\lb{velder}
V^{l_\pm}_{R\bk \a }=\partial \epsilon_{R}^{l_\pm}(\bk)/\partial k_\a.
\ee

In the $T\ra 0$ limit we can approximate the Fermi function with a $\d(\o)$ 
which selects only states at the Fermi level $\o =0$. By further assuming 
$\G_R^{l_\pm}$ to be small we can also approximate the spectral functions with a 
delta function and Eq.~\pref{eq:DC} reduces to 
\be
\lb{eq:DCT0}
\sigma^{l_\pm}_{\a}=\frac{ e^2}{N}\sum_{\bk} 
\frac{({V}_{R\bk_\a }^{l_\pm})^2}{\G_{R \bk}^{l_\pm}}  \ 
\delta (\epsilon_{R \bold{k}}^{l_\pm}).
\ee

\subsection{Analytical calculation}

To gain physical insight on the DC anisotropy and disentangle the effect of the 
velocity and scattering rate in Eq.~\pref{eq:DCT0}, we estimate analytically 
$\sigma_\alpha$. We approximate the h- and e-bands with perfectly nested 
parabolic bands and assume that the nematic order is small enough to allow one 
for a perturbative expansion of the renormalized energy $E_R^{l\pm}$. We use a 
symmetric nematic splitting around the isotropic value $\Sigma^l_0$ in the 
tetragonal phase. By expanding Eq.\pref{ER} at first order in the $\tau_3$ 
self-energy component we can estimate analytically for each pocket ${V}_{R\bk_\a 
}^{l_\pm}$ and $\Gamma^{l_\pm}_{R \bk}$ via Eq.s \pref{ReR}, \pref{GR} and 
\pref{velder}.

Deriving for example with respect $k_x$ the renormalized energy of the pocket 
$\Gamma_+$ we find 
\be
\lb{VG+_k}
V^{\G_+}_{R \bk x}= - \frac{k \cos \theta}{\, m^{\G_+}} 
+ 4  \, Re\Sigma^{\G_+}_3 \sin^2 \theta \frac{k \cos \theta}{k^2}. 
\ee
where $m^{\G_+}$ is the bare mass of the $\G_+$ pocket whose definition in terms 
of the Hamiltonian parameters is given in App.\,B. The first term on the r.h.s. 
of Eq.\pref{VG+_k}, is the $x$-component of the bare velocity, while the second 
term ${\cal O}(Re\Sigma^{\G}_3)$ is an additional contribution due to the 
orbital mixing induced by the nematic order as expected from the $(u,v)^l_R$ 
factors in Eq.\pref{vel}. To compute the $\bk$ integration in Eq.~\pref{eq:DCT0} 
we use the delta function and evaluate $V^{\G_+}_{R \bk x}$ at the renormalized 
FS. Notice that, in the nematic phase $k^{\G_+}_F (\theta)$ is no longer 
constant but gets deformed because of the anisotropic self-energy 
renormalization. This effect is also of order ${\cal O}(Re\Sigma^{\G}_3)$ and 
has to be taken into account. We estimate the change in the Fermi wave vector at 
the first order in the self-energy. Replacing the expression of 
$k^{\G_+}_F(\theta)$ into Eq.~\pref{VG+_k} we find 
\be
\lb{VG+_fin}
{V}^{\G_+}_{Rx}= {V_0}^{\Gamma_+}_x \bigg( 1 +  \cos 2\theta
\frac{Re\Sigma^{\G}_3}{2 \epsilon^{h}_0} - 4 \sin^2 \theta
\frac{Re\Sigma^{\G}_3}{2 \epsilon^{h}_0}  \bigg),
\ee
where ${V_0}^{\Gamma_+}_x= - {k_0}_{F x}^{\Gamma_+}/ m^{\G_+}$ and 
$\epsilon^{h}_0= \epsilon^{\G} + Re\Sigma_0^{\Gamma}$ are the velocity and the 
Fermi energy in the te\-tra\-go\-nal phase respectively. From 
Eq.~\pref{VG+_fin}, one sees that the bare Fermi velocity in the nematic phase 
has two contributions  ${\cal O}(Re\Sigma^{\G}_3)$ opposite in sign: the first 
one is due to the change in ${k}_{F}^{\Gamma_+}$, while the second one comes 
from the orbital mixing produced by the nematic order. Analogous calculation of 
the velocity contributions along $y$ for the $\Gamma_+$ as well as for the other 
pockets lead to similar expressions (see Eq.s~\pref{ii} in App.\,B) with the 
band velocity of the tetragonal phase renormalized by two additional 
contributions ${\cal O}(Re\Sigma^{l}_3)$ of opposite sign.
The scattering rate is analytically estimated from Eq.\pref{GR} using again the 
expansion of $E^{\G_+}_R$ at the first order in $\Sigma^{l}_3$
\be
\lb{GammaG+_fin}
\Gamma^{\G_+}_{R} (\theta) = \Gamma^{h}_{0} + cos 2\theta |Im\Sigma^{\G}_3|.
\ee
Here we separate the tetragonal phase scattering rate, $\Gamma^{h}_{0} =
\d \Gamma +  |Im\Sigma^{\G}_0|$, from the the angular-dependent correction due to the
nematic effect $\sim Im\Sigma^{\G}_3$.  
The explicit expression of the $\G_+$ pocket DC conductivity follows from 
Eq.~\pref{eq:DCT0} using Eq.s~\pref{VG+_fin}-\pref{GammaG+_fin}:
\bea
\s_{x}^{\G_+}&=&\s^h\bigg(1+ \frac{{Re}\Sigma_3^{\G}}{2\epsilon^h_0} - 
\frac{{Re} \Sigma_3^{\G}}{\epsilon^h_0} -
\frac{|{Im}\Sigma^\G_3|}{2\G^{h}_0}\bigg)\nn \\
&=&\s^h\bigg(1 - \frac{\Phi^h}{2\epsilon^h_0} + \frac{\Phi^h}{\epsilon^h_0} -
\frac{\D \G^h}{2\G^{h}_0}\bigg)
\lb{G+_xDC}
\eea
$\epsilon^{h}_0$, $\G^{h}_0$ and $\sigma^{h}= e^2\epsilon^{h}_0/(2 \pi
\hbar)\G_0^{h}$ are respectively the Fermi energy, the scattering rate and the
DC conductivity in the tetragonal phase. We also defined the real and imaginary
part of the nematic order parameter for the h-pockets ($\Phi^h, \Delta\Gamma^h$)
as
\bea
&\Phi{^h} \equiv \displaystyle\frac{Re\Sigma^{\G}_{xz}-Re\Sigma^{\G}_{yz}}{2}=
-Re \Sigma^{\G}_3 ,  \nn & \nn \\
\nn\\
& \Delta\Gamma{^h}\equiv|Im \Sigma^{\G}_3|, &
\lb{NOP_h}
\eea
taking also into account that stronger spin-fluctuation at $\bQ_X$ implies
${Re}\Sigma_3^{\G}<0$, so that now the nematic order parameters are all positive
defined.

Performing analogous calculations (see App.\,B) we derive the DC
conductivities along $x$ and $y$ for each pocket. For the h-pockets we find:
\bea
\s_{x/y}^{\G_+}&=&\s^h\bigg(1 \mp \frac{\Phi^h}{2\epsilon^h_0} \pm 
\frac{\Phi^h}{\epsilon^h_0}\mp \frac{\D \G^h}{2\G^{h}_0}\bigg), \nn \\ 
\s_{x/y}^{\G_-}&=&\s^h\bigg(1 \pm \frac{\Phi^h}{2\epsilon^h_0} \mp 
\frac{\Phi^h}{\epsilon^h_0} \pm \frac{\D \G^h}{2\G^{h}_0}\bigg).
\lb{holeDC}
\eea
In the absence of spin-orbit interaction the h-pockets have the same
$\epsilon_0^h$, so they also have the same conductivity $\s^h$ in the tetragonal
phase. Additional terms proportional to $\Phi^h$ and $\D \G^h$ arise in the
nematic phase and make the conductivity different for the two h-pocket. As
extensively discussed within the calculation of the velocity operator for the
$\G_+$ pocket in Eq.~\pref{VG+_fin}, the nematic order has two opposite effects
${\cal O}(\Phi^h)$ in the velocity and this is reflected into the pocket DC
conductivity anisotropy as one sees from Eq.s~\pref{holeDC}. The first
correction comes directly from the $k^{\G_\pm}_F$ changes due to the nematic FS
reconstruction, while the second one, opposite in sign, is due to the orbital
mixing. Notice that this last term also determine the overall sign of the
correction $\sim \Phi^h$ in each pocket.  
Due to the $xz/yz$ orbital arrangement of the $\G_{\pm}$ FS, the 
two h-pockets contribute with opposite sign to the conductivity anisotropy i.e. 
in Eq.~\pref{holeDC} we find the same sign of the nematic terms in the 
conductivity along $x$ of the $\G_+$ pocket and in the conductivity along $y$ of 
the $\G_-$ one. In particular the opposite sign of the contribution ${\cal O}(\D 
\G^h)$ giving negative/positive anisotropy for the $\G_{+/-}$ pocket is a direct 
consequence of the cold-spots physics, Fig.~\ref{fig-OSSF}, from where we 
can easily infer the sign of the anisotropic contribution for $\G_\pm$ having in 
mind that lower scattering implies a bigger conductivity.
By computing the h-DC conductivity anisotropy, $\Delta\sigma^{h_\pm}\equiv 
\sigma^{\G_\pm}_x-\sigma^{\G_\pm}_y$, we find: 
\bea
\Delta \sigma^{h+} &=& \s^h \left(\frac{\Phi^h}{\epsilon^h_0} -
\frac{\Delta \G^h}{\G^h_0} \right),
\nn \\
\Delta \sigma^{h-} &=& \s^h \left(-\frac{\Phi^h}{ \epsilon^h_0} +
\frac{\Delta \G^h}{\G^h_0} \right).
\lb{finalDC_h}
\eea

The DC conductivity components for the e-pocket at $X$ read
\bea
\s_{x}^{X}&=&\s^e\bigg(1 - \frac{ Re\Sigma^X_{yx}}{4 \epsilon^e_0} +
\frac{|Im\Sigma^X_{yz}|}{4 \G^e_0} \bigg), \nn \\
\s_{y}^{X}&=&\s^e\bigg(1 + \frac{3 Re\Sigma^X_{yx}}{4 \epsilon^e_0} - 
\frac{ |Im\Sigma^X_{yz}|}{4 \G^e_0}\bigg).
\lb{e_DC}
\eea
As already done in Eq.~\pref{holeDC1} we defined the nematic correction with
respect to the tetragonal $x/y$ DC conductivities. $\sigma_{x/y}^{e}$ are both
equivalent to $\sigma^{e}= e^2\epsilon^{e}_0/(2 \pi \hbar)\G_0^{e}$ since within
the parabolic band approximation we neglect the ellipticity of the e-pockets.
The same expressions of Eq.s~\pref{e_DC} hold for the $Y$ pocket once replaced
$\Sigma^X_{yz} \ra \Sigma^Y_{xz}$ and $k_x \ra k_y$. Thus also the $X/Y$ pockets
contribute with the opposite sign to the overall DC conductivity. By defining
the real and imaginary part of the e-pocket nematic order parameter $(\Phi^e,
\Delta\Gamma^e$):
\bea
\Phi{^e}\equiv \frac{Re\Sigma^X_{yz}-Re\Sigma^Y_{xz}}{2} ,
\quad  \quad  \Delta\Gamma{^e}=\frac{|Im \Sigma^X_{yz}|
-|Im \Sigma^Y_{xz}|}{2}\nonumber \\
\lb{NOP_e}
\eea
we can write the electronic DC conductivity anisotropy $\Delta\sigma^{e}\equiv 
\Delta \sigma^{X} + \Delta \sigma^{Y}$ as
\be
\lb{finalDC_e}
\Delta \sigma^e =  \s^e \left(-\frac{\Phi^e}{\epsilon^e_0}  
+ \frac{\Delta \G^e}{\G^e_0} \right)
\ee
Also for the e-pockets we find that the renormalized velocity and the scattering 
rate contribute with opposite sign to the DC conductivity anisotropy. The 
balance between the two effects is controlled by the nematic order parameters 
normalized to the Fermi energy and isotropic scattering rate, respectively, i.e. 
$\Phi^e/\epsilon^{e}_0$ vs $\Delta \G^{e}/\G^{e}_0$.

Summarizing, we computed analytically the anisotropy of the DC conductivity of
the various pockets using the parabolic-band approximation. We find for all the
pockets that the anisotropy is given by a contribution ${\cal O}(Re \Sigma^{l})$
and another ${\cal O}(Im \Sigma^{l})$, opposite in sign with respect to each
other, whose relevance is controlled by the values of
$\Phi^{h/e}/\epsilon^{h/e}_0$ vs $\Delta \G^{h/e}/\G^{h/e}_0$. 
Summing up the h-and e-pockets $\Delta\s^{h/e}$ we find that the sign of the
anisotropy of the total DC conductivity depends on which pocket contributes more
to the total conductivity and on which effect, among the scattering rate and
velocity renormalization, dominates. Within the perfectly nested parabolic band
approximation, in which the two h-pockets are equivalent, their anisotropic
contributions, Eq.s~\pref{finalDC_h}, are opposite in sign and cancel out, so
that the DC anisotropy is determined only by the e-pockets. In this situation
the overall sign of $\Delta\sigma$ depends on which effect dominates in
$\Delta\sigma^e$, Eq.~\pref{finalDC_e} i.e. the anisotropy of the velocity
or the one of the scattering rate.

In real IBS systems, however, we need to account for the presence of the 
spin-orbit interaction that splits the h-pockets at $\G$ and mixes their orbital 
content at the FS already in the tetragonal phase. Moreover, the parabolic band 
approximation is particularly inaccurate for the e-pockets that are strongly 
elliptical in all IBS. Furthermore, especially for FeSe, the nematic self-energy 
components $\Sigma^l_3$ are not small\cite{Fanfarilloprb16}, thus the expansion 
of the renormalized energy in $\Sigma^l_3$  performed above is not longer 
justified. For realistic cases then, we cannot use the analytical expressions, 
Eq.s~\pref{VG+_fin}-\pref{GammaG+_fin}, and we need to compute the DC 
conductivity from Eq.\pref{eq:DCT0} using a numerical estimate of the velocity 
and scattering rate from Eq.s~\pref{ER}-\pref{GR} and \pref{velder}. 

\subsection{Beyond the analytical approach} \label{numeric} We perform a 
numerical estimate of the conductivity anisotropy using realistic parameters for 
122 and FeSe systems in order to assess the limits of validity of the analytical 
expressions Eq.s~\pref{finalDC_h} and \pref{finalDC_e} and qualitatively discuss 
our results in the context of the experimental outcomes found for 122 pnictides 
and FeSe. We assume for both 122 and FeSe equivalent band-structure parameters 
that results in the tetragonal FS shown in Fig.\ref{fig-OSSF}. The FS topology 
of FeSe with just the outer h-pocket crossing the Fermi level at $\G$ already in 
the tetragonal phase is achieved in the calculation using a larger value of the 
spin-orbit interaction as well as a larger values of the real part of the 
self-energy renormalizations in agreement with previous 
analysis\cite{Fanfarilloprb16}. The numerical values of the parameters used in 
the following are detailed in App.\,C.

\paragraph{122 pnictides} 
\begin{figure}[tbh] 
{\includegraphics[width=\linewidth]{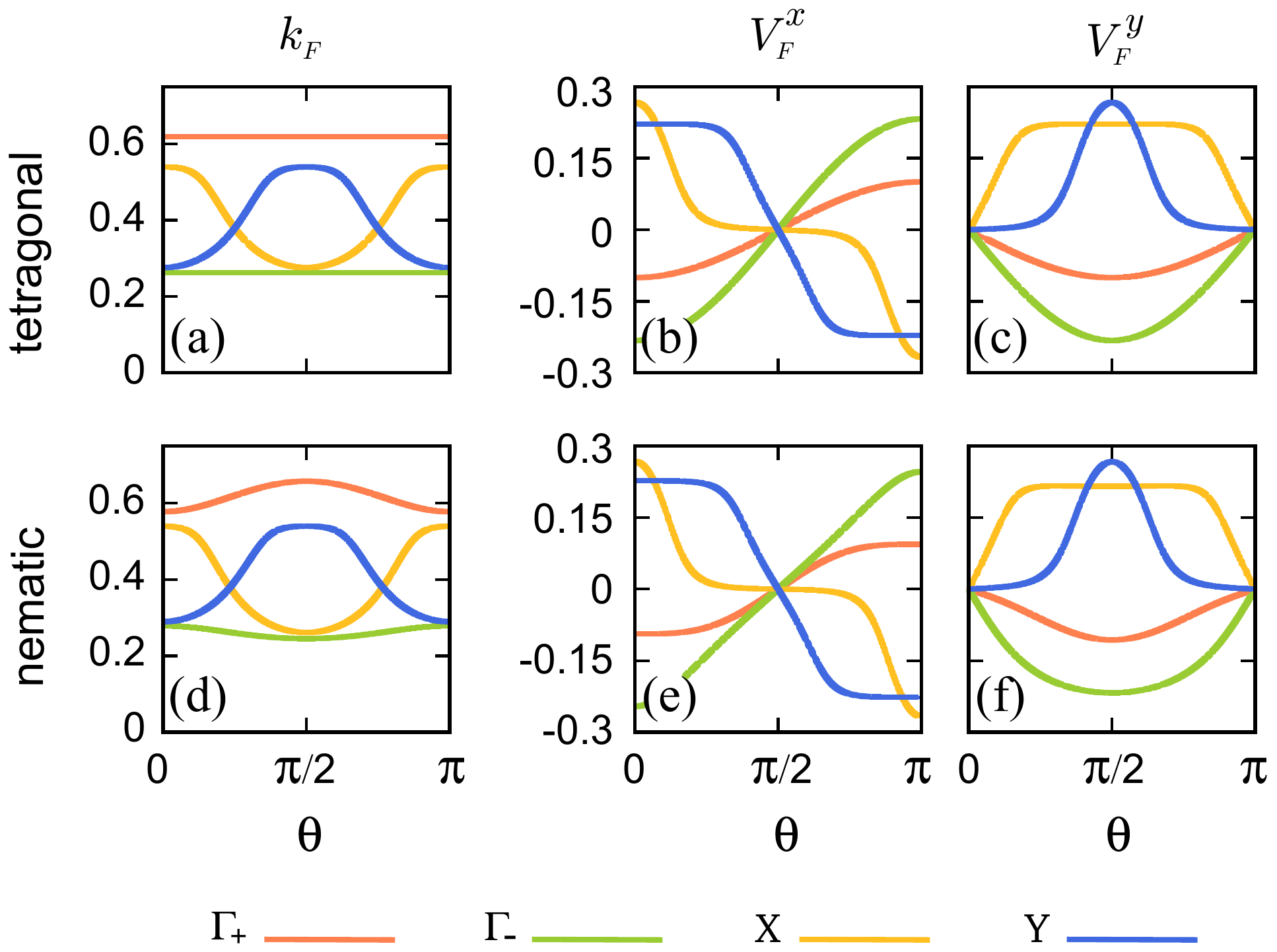}} \caption{Numerical 
computation of FS wavevectors and velocity components for 122 system parameters 
in the tetragonal and nematic phase. $\Phi_h=\Phi_e= 4$ meV, the spin-orbit 
interaction is $5$ meV, other band parameters are detailed in App.\,C. The $k_F$ 
are measured in units $1/a \sim 0.375$ \AA, where $a=a_{FeFe}$ is the lattice 
constant of the 1-Fe unit cell. The velocities are in eV.} 
\label{fig:kF_vs_vf_122}
\end{figure}
\begin{figure}[tbh] 
\vspace{-.5cm}
{\includegraphics[width=0.9\linewidth]{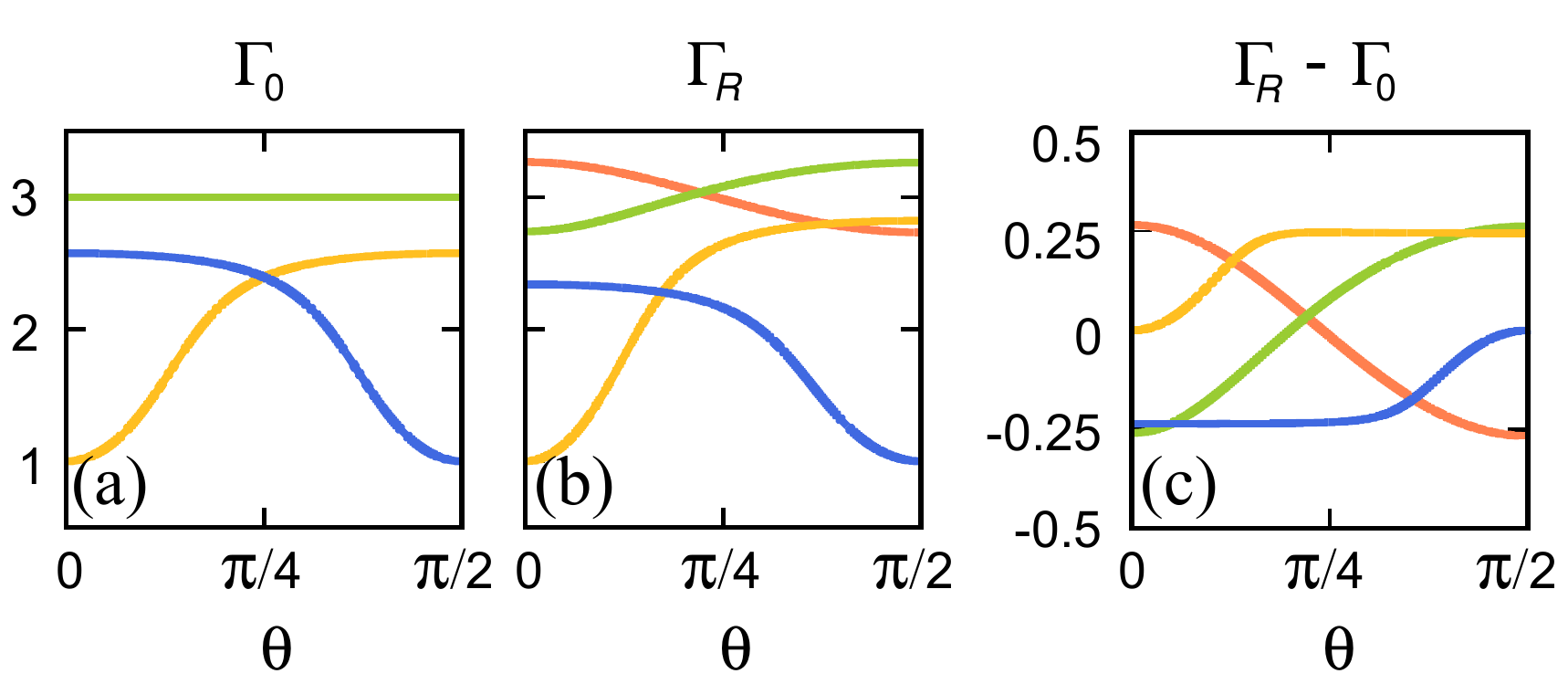}}
\caption{Renormalized scattering rate (in meV) computed using realistic 122 system
parameters. Here $\G^h_0= 3$ meV and $\G^e_0= 2$ meV. We fix $\D \G^h$ and $\D
\G^e$ considering the imaginary part of the self-energy for each pocket changing
proportionally to the real part in the nematic phase (App.\,C).}
\label{fig:scatt_122}
\vspace{-0.2cm}
\end{figure}
\begin{figure}[tbh] 
\includegraphics[width=\linewidth]{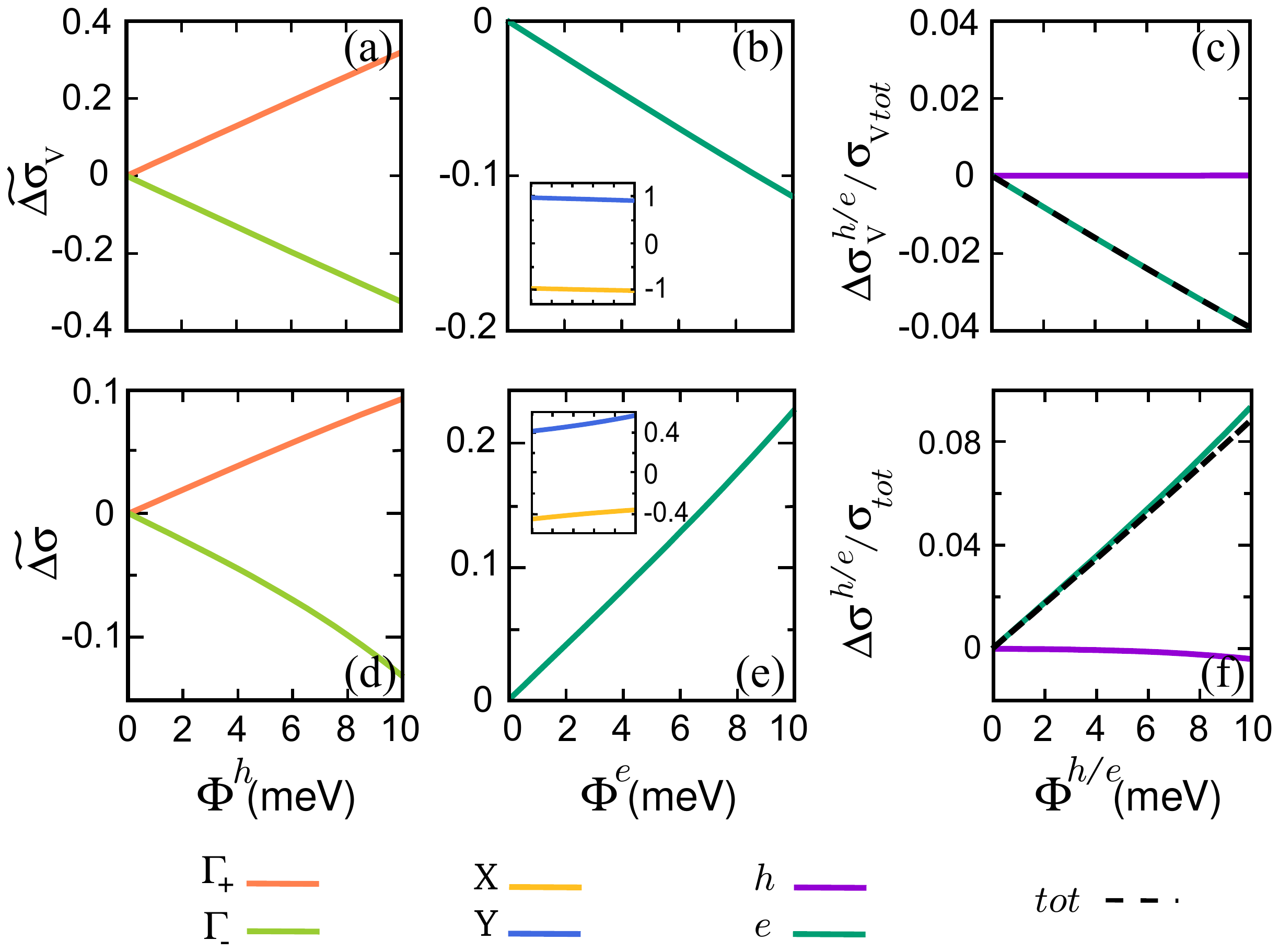} \caption{Numerical
computation of the velocity contribution to the DC conductivity anisotropy
$\D\sigma_V$ and of the total DC conductivity anisotropy $\D \s$ for realistic
parameter for 122. In panels a-b we renormalized the pocket contributions to
their value in the tetragonal phase i.e. $\D \sigma_V^{l_\pm}(\Phi^{h/e}=0)$ and
analogously in panels d-e. In c and f panels instead we renormalize the h- and
e-contributions to the total tetragonal values, i.e. $ \sigma_{Vtot}=
\sigma_V(\Phi^{h/e}=0)$ and $\s_{tot}=\s(\Phi^{h/e}=0)$.} 
\label{fig:DC_122}
\vspace{-0.5cm}
\end{figure}
In Fig.~\ref{fig:kF_vs_vf_122} we show for each pockets the FS wavevectors and 
velocities along $x/y$. To better appreciate the changes induced by the nematic 
order we plot in the first row the results for the tetragonal phase and in the 
second ones the results obtained in the nematic phase assuming $\Phi^h =\Phi^e= 
4$ meV. The h-pockets, circular in the tetragonal phase, are weakly deformed in 
the nematic phase due to the small nematic order that also makes the $X/Y$ 
pockets slightly different (a-d panel). The changes in the velocities for the 
$\G_{\pm}$ pockets appear to be quite small and do not follow monotonously the 
renormalization of the Fermi vectors as one could have expected (e-f panels). 
This is in agreement with the analytical calculation outlined above, where we 
found that the renormalization in the velocities due to the orbital mixing and 
the one coming from the Fermi vector renormalization are opposite in sign, 
reducing the overall anisotropic effect on the velocity (see Eq.\pref{VG+_fin}). 
Due to the ellipticity of the FS, the e-pockets have anisotropic velocities 
already in the tetragonal phase (b-c panels) with the $X/Y$ pockets showing 
larger velocity along $y/x$. No qualitatively changes are visible in the nematic 
phase (e-f panels). 
In Fig.~\ref{fig:scatt_122} we show for each pocket the scattering rate obtained
from Eq.~\pref{GR}. In the three panels we show the tetragonal value, $\G_0$,
the nematic one $\G_R$ and their difference. We find again a good agreement between
the analytical calculations and the numerical results for the h-pockets. As in
Eq.~\pref{GammaG+_fin} the angular dependence of the correction
$\G^{\G_\pm}_R-\G^{\G_\pm}_0 $ goes almost as a $\cos 2\theta$, even if the weak
ellipticity of the h-FS induced by the nematic order causes minor deviations,
e.g. the correction vanishes for the $\G_{+}/\G_-$ slightly before/after
$\pi/4$. No renormalizations are found along $x/y$ for the $X/Y$ pockets since,
within our model, no scattering is allowed in the $xy$ channel
(Fig.\ref{fig-OSSF}). The location of the so-called cold spots i.e. the position
of the minima of the scattering rate for both h- and e-pockets, does not change
once a realistic FS are considered and corresponds to the ones shown in
Fig.\ref{fig-OSSF}.

We can disentangle the effect of the velocity and of the scattering rate on the
DC anisotropy by computing Eq.\pref{eq:DCT0} using a constant scattering rate.
This result just account for the anisotropic effects coming from the velocity 
so we will refer to it as $\D\sigma_V$. In Fig. \ref{fig:DC_122} we show
for each pockets $\D \sigma_V$ (a-c panels) and the complete conductivity
anisotropy $\D\s$ (d-f panels) as a function of $\Phi^{h/e}$.
To easily compare the results of the numerics with the analytical estimate of 
Eq.s~\pref{finalDC_h} and \pref{finalDC_e} we renormalized the $h_\pm/e$ pocket 
anisotropy in panels a-b and d-e to their value in the tetragonal phase. In 
panel c and f we renormalize instead the h- and e-anisotropy to the total values 
of $\sigma_V$ and $\s$ obtained summing all the pockets contributions in the 
tetragonal phase. From the analysis of $\D\sigma_V$ we find that the sign of the 
anisotropic contribution proportional to $\Phi^{h/e}$ found in 
Eq.s~\pref{finalDC_h} and \pref{finalDC_e} is robust, with the $\G_{+/-}$ and 
the $Y/X$ pockets contributing with positive/negative terms to the 
DC-conductivity anisotropy (see panels a and b and inset of b). The h-pockets 
anisotropy due to the velocity, panels a-c, are opposite in sign and grows as 
$\Phi^h/\epsilon_0^h$ in agreement with the analytical expectation. Even if the 
$\G_{\pm}$ are not longer equivalent due to a small spin-orbit interaction their 
anisotropic contributions almost cancel out so that the negative anisotropy of 
the e-pocket is the one that determines the final results. 
Once the effect of the scattering rate is included in the calculation we see in 
Fig.\ref{fig:DC_122} d a reduction of the conductivity anisotropy for the 
h-pocket that however still sum up to an anisotropic conductivity terms close to 
zero (panel f). In contrast a change of sign in the overall electronic term is 
observed due to the larger positive contribution $\D\sigma^Y$ of the $Y$ pocket 
once that the anisotropic scattering rate is correctly taken into account. For 
the set of parameters used, thus, we find a final $\D\sigma>0$. The result comes 
from the change in the relative weight of the contribution of the $X$ and $Y$ 
pockets in the e-term due to the different scattering rate $\G^{X/Y}_R$. The 
final outcome is thus particularly sensitive to the $\G^e_0$ and $\D \G^e$ used 
and could be strongly affected by any mechanism (temperature, disorder, 
interactions, ect.) affecting their absolute values. 

\begin{figure}[tbh]
\centering
\includegraphics[width=\linewidth]{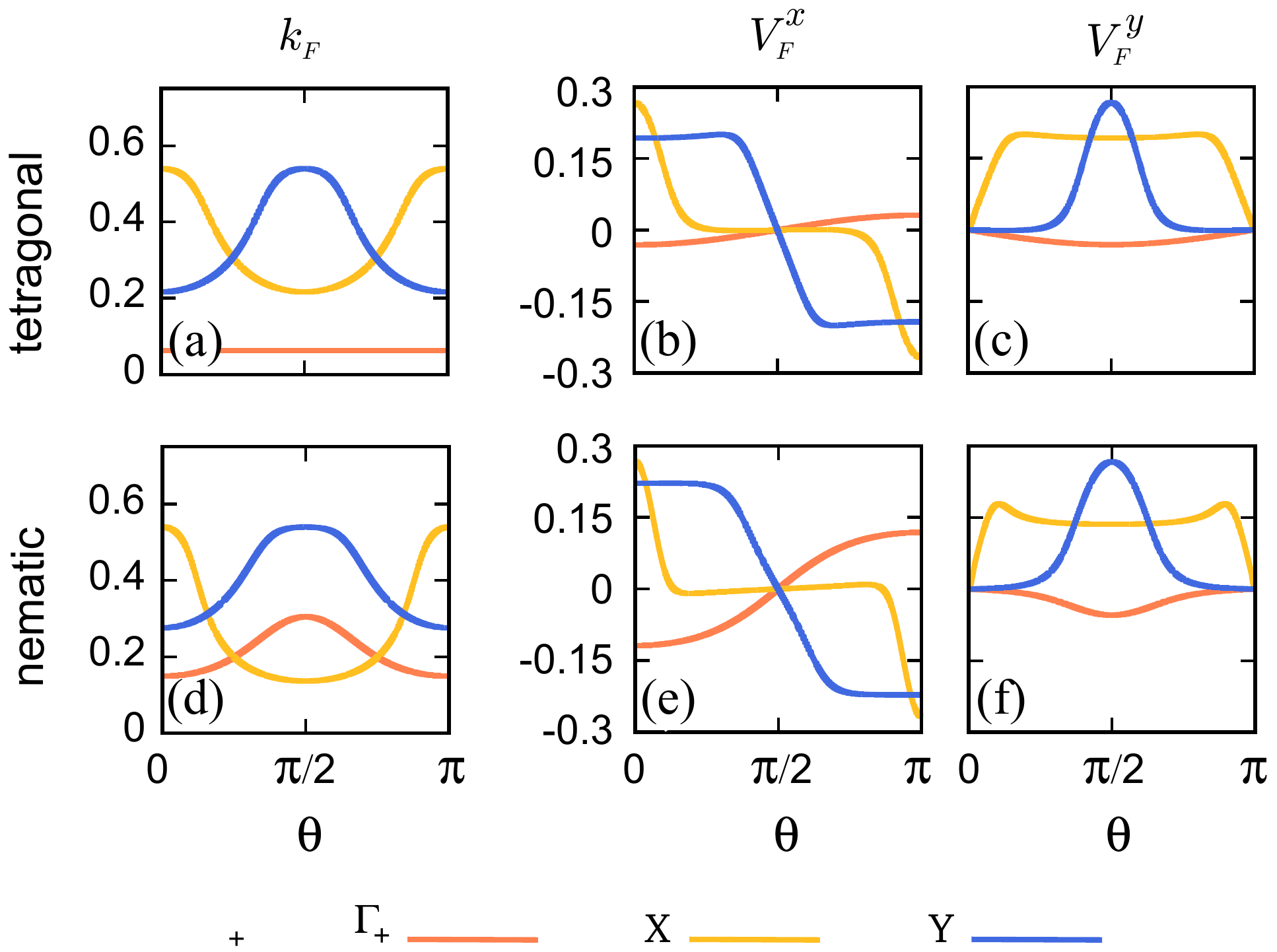} 
\caption{Numerical computation of the tetragonal and nematic FS and Fermi
velocity components for FeSe parameters. $\Phi_h=\Phi_e= 15$ meV, the spin-orbit
interaction is $20$ meV, the other band parameters are detailed in App.\,C.} 
\label{fig:kF_vs_vf_FeSe}
{\includegraphics[width=0.9\linewidth]{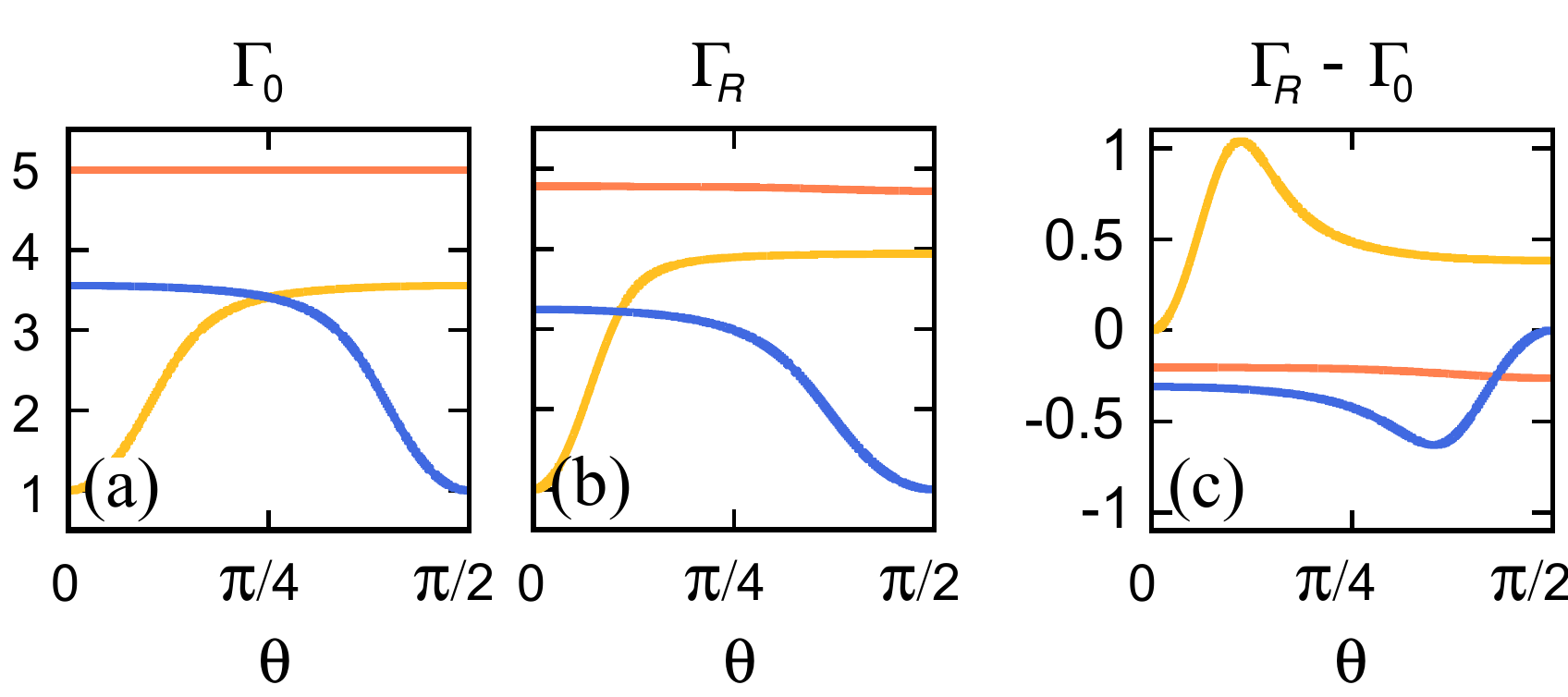}}
\caption{Renormalized scattering rate for FeSe. $\G^h_0= 5$ meV and $\G^e_0=
2.5$ meV. $\D \G^h$ and $\D \G^e$ are considered as proportional to the nematic
variation of the real parts (App.\,C)} 
\label{fig:scatt_FeSe}
\end{figure}

\paragraph{FeSe} 
We repeat the numerical analysis considering the case of FeSe. In
Fig.~\ref{fig:kF_vs_vf_FeSe} we show the pockets FS wavevectors and velocities
both in the tetragonal and in the nematic phase assuming $\Phi^h =\Phi^e= 15$
meV. 
\begin{figure}[tbh] 
\centering
\subfigure{\includegraphics[width=\linewidth]{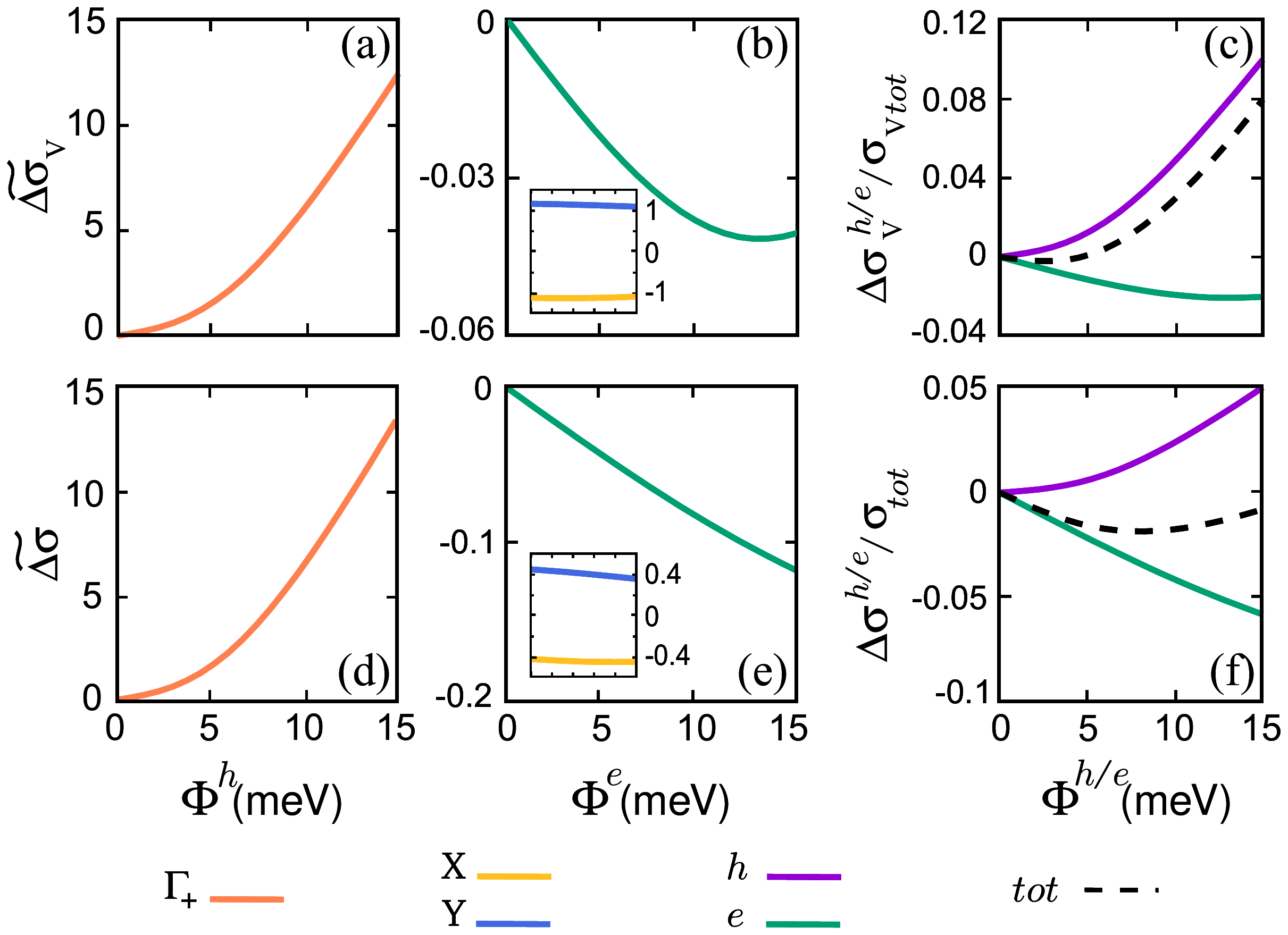}} 
\caption{Numerical computation of the velocity contribution to the DC 
conductivity anisotropy and of the total DC conductivity anisotropy for 
realistic parameter for FeSe. All the pockets show consistent deviations of the 
DC anisotropy with respect the analytical expectations Eq.s~\pref{finalDC_h} and 
\pref{finalDC_e}. } 
\label{fig:DC_FeSe}
\vspace{-0.6cm}
\end{figure}

With respect to the previous case, here we clearly see that the outer h-pocket, 
the only one crossing the Fermi level, becomes strongly elliptical in the 
nematic phase due to the large nematic order which also makes the $X/Y$ pockets 
consistently different in size (a-d panel). The changes in the velocities (e-f 
panels) are similar to the ones observed for the 122 case but quantitatively 
more pronounced here due to the larger value of the nematic order parameters. 
The scattering rates for all the pockets are shown in Fig. \ref{fig:scatt_FeSe}.
For all the pockets we find a clear deviation of the renormalized scattering
rate from the analytical estimate. In particular, the angular dependence of the
$\G_+$ scattering rate is very weak and does not resemble the $\cos 2 \theta$
predicted by Eq.\pref{GammaG+_fin}. This is a consequence of the FS nematic
reconstruction of FeSe. In fact, the nematic order not only makes the $\G_{+}$
pocket elliptical but also affect its orbital content that becomes almost
completely $xz$ at the Fermi level\cite{Fanfarilloprb16, Rhodesprb18,
Zhouarxiv18}. As a consequence, the cold spots of the outer pocket shown in
Fig.\ref{fig-OSSF} do not represent anymore a minimum of the scattering since
the $\G_+$ FS is mostly $xz$ also at $\theta=0$. 

We study also in this case for each pocket the behavior of $\D\sigma_V$ and $\D \s$
as a function of $\Phi^{h/e}$, Fig.~\ref{fig:DC_FeSe}. We use the same
renormalizations used in Fig.~\ref{fig:DC_122}. The analysis of the velocity
contribution reveals that the sign of the $\Phi^{h/e}$ terms of
Eq.s~\pref{finalDC_h},\pref{finalDC_e} is robust also in this case. We are no
longer in the perturbative regime as one can see from the non-linear behavior of
$\D\sigma_V^{h_+}$, shown in panel a, where the $\G_+$ pocket anisotropy
contribution grows much faster than what expected from the linear dependence in
Eq.~\pref{finalDC_h}. The final anisotropy of $\D \sigma_V$ is the result of the
competition between the h- and e-terms. 
The inclusion of the scattering rate in the calculation strongly affect the 
e-pockets contribution while leaving  $\D \sigma^{h_+}$ almost unchanged. As a 
matter of fact, the scattering rate of the outer h-pocket $\G^{\G_+}_R$ is 
almost isotropic, see Fig.\ref{fig:scatt_FeSe}, so that the anisotropic velocity 
is the only factor which contributes to the DC anisotropy of the $\G_+$ pocket. 
The final result for the DC conductivity strongly depends on the set of 
$\G^{h/e}_0$ and $\D \G^{h/e}$ used. In Fig.\ref{fig:DC_FeSe} d-f, we show a 
case in which the inclusion of the scattering rate enhances the relative weight 
of the e-contribution with respect the h-term, so that overall the negative 
anisotropy of the e-part, due to the $X$ pocket, determines the final results 
shown in Fig.\ref{fig:DC_FeSe} f. 

\paragraph{High-energy renormalization and nesting}

An effect neglected in the above calculation is the renormalization of the 
quasiparticle due to local electronic interactions. It is well established that 
in IBS the high-energy renormalizations of the quasiparticle $Z_{orb}$ coming 
from local interactions are quite strong and orbital-dependent. This effect, 
usually named orbital-selective renormalization in the 
literature,\cite{Davis2017} should not be confused with the effects discussed in 
the present work, where the orbital selectivity refers to the spin fluctuations, 
which affect the various orbital in a different way in the nematic phase. The 
high-energy renormalizations have noticeable effects on the optical conductivity 
in the tetragonal phase of IBS, as discussed in 
Ref.~[\onlinecite{Calderonprb14}] and should be included in the above 
calculation. 
We then repeted the numerical analysis including phenomenologically the orbital
renormalizations into the coherence factors $({u/v})^l_{R}$ entering in
Eq.~\pref{vel}. This allows us to estimate the effects of the high-energy
renormalization on the velocity contributions to the DC conductivity $\D
\sigma_V$. As expected, the inclusion of a severe reduction of the coherence of
the $xy$ orbital ($Z_{xy}\sim 0.3$), that is the most correlated orbital in all
IBS, leads to the suppression of the $V^{x/y}_{X/Y}$ contributions enhancing the
DC anisotropy in the e-pockets. Moreover, the small differentiation ($\sim
10\%$) of the quasiparticle masses for the $xz/yz$ orbitals in the nematic
phase\cite{Fanfarilloprb17} contributes to enhance the differentiation of the
$\G_\pm$ and $X/Y$ pockets. However, the sign of the velocity contribution to
the DC anisotropy is robust within the set of $Z_{orb}$ considered. The
quasiparticle renormalizations affect also the conductivity via the renormalized
scattering rate, however their relevance strongly depends on the set of
parameters used, whose analysis goes beyond the scope of the present work.

Finally, it is worth noting that the analysis presented here,  as well as the 
one carried out previously\cite{Fanfarilloprb16} within a momentum-independent 
Eliashberg scheme, does not  include the physics of the band-nesting effects, 
which  are the only ones to determine the location of the hot-spots in the 
band-based description \cite{Fernandesprl10,prozorovnatcomm13}. In particular 
the ellipticity of the $X/Y$ pockets suggests that the scattering rate is maximum 
at the location where the nesting with the h-pockets is realized.  Within the 
OSSF both the orbital character and the degree of band nesting will contribute 
to the hot-spot location. Recent multiorbital RPA calculation in the 
paramagnetic state support the idea that the dominant effect in determining the 
scattering rate is still the orbital character of the FS\cite{Hirschfeldprb11, 
KorshunovarXiv17}. How these results evolve below $T_S$ within an 
orbital-selective spin-nematic scenario is still an open question which 
certainly deserves further investigation.

\subsection{Comparison with experiments}
In the previous section we computed numerically the DC anisotropy for realistic 
parameters of 122 pnictides and FeSe. The values of the band structure 
parameters and self-energies used in the calculations quantitatively 
reproduce the main features of the FS, including the FS shrinking and the orbital 
FS reconstruction experimentally observed in the nematic phase of 122 and FeSe. 
In 122, where the nematic order parameters $\Phi^{h/e}$ are small, the h-pockets 
contribution to the DC conductivity anisotropy is well approximated by the 
analytical estimate Eq.~\pref{finalDC_h}, while we observe consistent deviations 
 in FeSe. Nonetheless, for both 122 and FeSe systems the sign of the anisotropic 
contribution coming from the renormalized velocity, $\D \sigma_V$, is 
robust. In both cases considered in sec.~\ref{numeric} we managed to match 
the experimental result, $\Delta \sigma_{DC}(FeSe)<0$ and $\Delta 
\sigma_{DC}(122)>0$, once the renormalization of the scattering rate is included 
in the calculation. As already mentioned, the final result is still somehow
sensitive to the set of parameters used. Thus, in this last section we discuss in 
general which are the possibilities to match the experimental results regardless 
the precise choice of parameters used in sec.~\ref{numeric}.

Concerning 122 systems, as long as the h-pockets contributions to the DC 
anisotropy cancel out, the final result is controlled by the e-pockets. Since 
they have a strong elliptical deformation,  their overall contribution to the DC 
anisotropy cannot be predicted from the analytical result Eq.\pref{finalDC_e}, 
and the final outcome depends on the relative weight of the $X$ and $Y$ pockets 
and on the relevance of the scattering-rate anisotropy over the contribution $\D 
\s^e_V$. Even in doped 122 compounds the h-pocket contributions cancellation 
still occurs since the relatively small value of the spin-orbit splitting at 
$\G$ guarantees that the Fermi energy is the same for the h-pockets. However, 
doping changes both the size of the pockets and the degree of nesting between h- 
and e-pockets. Both effects contribute to change the relative weight of the 
$X/Y$ e-pockets as well as the balance between the velocity vs scattering-rate 
anisotropic contributions and can be at the origin of the different sign of 
$\D\s_{DC}$ experimentally observed between the h- and e-doped side of 122 phase 
diagram. 

For what concerns FeSe, the presence of a single h-pocket and its strong 
orbital reconstruction lead to rather different physics. In particular, since 
the nematic FS reconstruction makes the whole $\G_+$ FS mostly $xz$ even at 
$\theta = 0$, the expected anisotropy of the renormalized scattering rate is 
absent, see  Fig.~\ref{fig:scatt_FeSe}. As a consequence the $\G_+$ anisotropic 
contribution is more likely controlled by the velocity anisotropy. This result 
should be contrasted with the outcomes of Ref.~[\onlinecite{Kontaniprb17}], 
where the difference between 122 and FeSe is fully ascribed to a different 
behavior of the scattering-rate anisotropy in the two compounds. In our picture 
the FeSe DC-conductivity anisotropy emerges from a subtle interplay between the 
competing effects coming from the scattering rate and the velocity, as it has 
been recently suggested by optical conductivity experiments in 
FeSe\cite{Degiorgiprb18}. It is worth noting that recent ARPES 
experiments reveal a strong $k_z$ dependence of the orbital composition of the 
$\G_+$ FS\cite{BorisenkoarXiv18}, with the FS a $k_z=\pi$ recovering 
$yz$-character at $\theta=0$. As a consequence, also the scattering rate 
anisotropy on $\G_+$ is expected to be larger at $k_z=\pi$ and its 
effect on the DC-conductivity anisotropy can possibly compete with the velocity 
term at this $k_z$. This observation calls for a more complete analysis of the 
DC anisotropy involving also the $k_z$-dependence of the FS.

\section{Conclusions}

In conclusion we computed the DC-conductivity anisotropy in the nematic phase of
IBS using the orbital-selective spin-nematic scenario that accounts for the
orbital content of the FS\cite{Fanfarilloprb15, Fanfarillo2018}. In this
scenario, the DC anisotropy of the nematic phase of IBS depends on
the scattering rate and velocity renormalizations due to self-energy corrections.
Both scattering rate and velocity are affected by the FS nematic reconstruction.
The scattering rate is strongly affected by the orbital content of the FS, and
the location of its minima on the FS is found in correspondence of the less
renormalized orbitals giving rise to cold spots. The velocity renormalization is
sensitive both to the orbital mixing and to the shrinking of the FS induced by
the nematic order, with the former effect dominating over the latter. Due to this
effect we find the unexpected result that the conductivity increases in the
direction in which the self-energy is larger and the shrinking is stronger. 
For both h- and e-carriers the contribution of the velocity to the DC anisotropy 
is opposite in sign to the one of the scattering rate. This is in agreement with 
recent optical conductivity experiment in FeSe\cite{Degiorgiprb18} where it is 
shown that scattering rate and velocity contribute to the conductivity 
anisotropy with opposite signs. Our results naturally follows from the 
spin-orbital entanglement implicit in the OSSF model and are new results in 
contrast to the band spin-nematic scenario\cite{schmalianprb12, 
fernandesnatphys14}. In particular we demonstrated that the usual expectation of 
anisotropic magnetic fluctuations giving rise only to an anisotropy in the 
inelastic scattering rate\cite{Fernandesprl10} is not longer valid once the 
orbital degree of freedom is taken into account in the theoretical description.

We performed numerical calculation for representative parameters for 122 
pnictides and FeSe. We verified that for 122 system the analytical estimate 
represents a good approximation of the numerical with the overall h-pockets 
contribution vanishing even once a finite spin-orbit splitting at $\G$ is 
considered. Numerical results for FeSe instead deviate from the analytical 
expectations due to the huge nematic FS reconstruction. We also discuss how the 
conductivity anisotropy depends on the system parameters. It can be dominated by 
either electron or hole pocket conductivity and depends on  ellipticity and 
high-energy renormalizations. The OSSF scenario provides then a suitable 
framework where the same mechanism due to orbital-spin interplay can reconcile 
the experimental observations reported in different families of iron-based 
superconductors.

\section*{ACKNOWLEDGMENTS}
We acknowledge L. Degiorgi for useful discussions. L.B. acknowledges financial 
support by Italian MAECI under the collaborative Italia-India project 
SuperTop-PGR04879. B.V. acknowledges funding from MINECO (Spain) via Grants 
No.FIS2014-53219-P and Fundaci\'on Ram\'on Areces. L.B., L.F. and B.V 
acknowledge the cost action Nanocohybri CA16218.

\section*{\bf{APPENDIX A: Green's function and perfectly nested parabolic band 
approximation.}} 

In this appendix we calculate the dressed Green's function to be used in the DC 
conductivity. We also detail the perfect nested parabolic band approximation to 
perform the analytical calculation of the conductivity.

\subsection*{Green's function}
The bare Hamiltonian consists in a 4-pocket model with two h-pockets at $\G$, 
$\G_\pm$ and two e-pockets at $X$ and $Y$ adapted from the low-energy model 
considered in Ref.\,[\onlinecite{Vafekprb13}]. Each pocket is described using a 
spinor representation in the pseudo-orbital space. 
\be 
H_0^l=\sum_{\bk,\s} 
\psi^{\dagger l}_{\bk\s} \hat H_{0 \bk}^l \psi^l_{\bk\s},
\ee
where $l=\G,X,Y$ and the spinors are defined as: $\psi^{\Gamma}_{\bk 
\s}=(c^{yz}_{\bk,\s},c^{xz}_{\bk \s})$ and $\psi^{X/Y}_{\bk \s}=(c^{yz/xz}_{\bk 
\s},c^{xy}_{\bk \s})$. The matrix $\hat H_{0 \bk}^l $ has the general form
\be
\lb{a}
\hat H_{0 \bk}^l=
h_0^l\red{\hat{\t_0}}+\vec{h}^l\cdot\red{\vec{\hat{\t}}} = 
\begin{pmatrix}
h_0^l + h_3^l \ \ & \ \ h_1^l - ih_2^l \\
 h_1^l + ih_2^l \ \ & \ \ h_0^l - h_3^l \\
\end{pmatrix}
\ee
with $\hat{\t}$ matrices representing the pseudo-orbital spin. The $h^\G$ 
components read as
\bea
h_0^{\G} &=& \epsilon\red{^\G} -a\red{^\G} \bk^2, \nn \\
h_1^{\G} &=& -2 b\red{^\G} k_x k_y, \nn \\
h_3^{\G} &=& b\red{^\G} (k_x^2 -k_y^2),
\lb{hG}
\eea
and for the $X$ pocket,
\bea
h_0^{X} &=& (h\red{^{yz}}+ h^{xy})/2 \nn \\
h_2^{X} &=& v k_y \nn\\
h_3^{X} &=& (h\red{^{yz}}- h^{xy})/2 - b (k_x^2 -k_y^2) 
\lb{hX}
\eea
where $h\red{^{yz}} = -\epsilon\red{^{yz}} + a\red{^{yz}} \bk^2$ and 
$h\red{^{xy}} = - \epsilon\red{^{xy}} + a\red{^{xy}} \bk^2$. Ana\-lo\-gous 
expressions hold for the $Y$ pocket provided that one exchange $k_x$ by $k_y$. 
Rotating the Hamiltonian $\hat H_{0 \bk}^l $  into the band basis we have $\hat 
H_{0 \bk}^l = \hat {\cal U}\red{^{l^{-1}}} \hat {\Lambda}^l \hat {\cal U}^l $ 
where $\hat {\Lambda}\red{^l} = diag (E^{l +}_{\bk}, E^{l -}_{\bk})$ is the 
eigenvalue matrix and $E^{l \pm}_{\bk}$ is given by 
\be 
\lb{b} 
E^{l \pm}_{\bk} = h_0^l\pm h^l = h_0^l\pm \sqrt{(h^l_{1})^2+(h^l_{2})^2 + 
(h^l_{3})^2}, 
\ee 
The fermionic band operators $c^{l\red{\pm}}$ from $H_0^l=\sum_{\bk,\s} E^{l 
\pm}_{\bk} {c}^{\dagger l\red{\pm}}_{\bk \s} c^{l\red{\pm}}_{\bk \s}$ are 
obtained rotating the orbital spinors via $\hat {\cal U}^l $. The unitary $\hat 
{\cal U}^l $ matrix has the common form: 
\be
\lb{dd}
 \hat {\cal U}^l  = \begin{pmatrix}
u^l & -v^l \\
v^{*l} & u^{*l}\\
\end{pmatrix}
\ee
where we drop the momentum and spin indices for simplicity. Explicitly for the 
h-pockets at $\Gamma$, $c^{\G\red{\pm}} \equiv h^{\red{\pm}}$ we have 
\be
\lb{d}
\begin{pmatrix}
h\red{^+} \\
h\red{^-} \\
\end{pmatrix} = 
\begin{pmatrix}
u\red{^{\Gamma}} &-v\red{^{\Gamma}}\\
v\red{^{*\Gamma}} & u\red{^{*\Gamma}}\\
\end{pmatrix}
\begin{pmatrix}
 c\red{^{yz}} \\
 c\red{^{xz}} \\
\end{pmatrix} =\red{{\hat{\cal U}}^\G}
\begin{pmatrix}
 c\red{^{yz}} \\
 c\red{^{xz}} \\
\end{pmatrix}
\ee
Analogous expressions hold for the $X/Y$ e-pockets fermionic operators 
$c^{X/Y\red{\pm}} \equiv e\red{^{X/Y \pm}}$, provided that the corresponding 
orbital spinors $\psi^{X/Y}=(c\red{^{yz/xz}}, c\red{^{xy}})$ are used. 
The coherence factors are:
\bea
\lb{c}
u^l &=&\frac{1}{\sqrt{2}} \sqrt{1+\frac{h_{3}^l}{h^l}}, \nn \\
v^{*l} &=&\frac{1}{\sqrt{2}}\frac{h_{1}^l + ih_{2}^l}{\sqrt{(h_{1}^l)^2 
+ (h_{2}^l)^2}} \sqrt{1-\frac{h_{3}^l}{h^l}}.
\eea
Knowing the dispersion relation and the coherence factors inside the rotation 
matrix $\red{\hat{\cal U}^l}$ we can obtain the bare pocket Green's functions. 
Since $\hat G_{0}\red{^{l^{-1}}}=\o\red{\hat{\mathbbm{1}}} -\hat H_{0 \bk}^l$ 
simply $\hat G_{0}\red{^{l^{-1}}}=\red{{\hat{\cal 
U}}^{l^{-1}}}(\o\red{\hat{\mathbbm{1}}} -\red{\hat{\Lambda}}^{l}){\red{\hat{\cal 
U}}}^l$ with $\red{\hat{\Lambda}}^{l} = diag (E^{l_+} , E^{l_-} ).$ 

We now turn our attention to the OSSF self-energy. The low-energy interacting 
Hamiltonian describing the spin exchange between h- and e-pockets is given in 
the main text Eq.~\pref{hint_low}. The corresponding Dyson equation for each 
pocket is: 
\be
\lb{eq:Dyson}
\red{\hat G^{l^{-1}}(\bk,\o)}=\red{\hat G_{0}^{l^{-1}}(\bk,\o)}-\hat \Sigma^l(\o),
\ee
where  $\hat G_{0}\red{^{l^{-1}}}(\bk,\o)=\o\red{\hat{\mathbbm{1}}} -\hat H_{0 
\bk}^l$ and $\hat\Sigma^l(\o)=\Sigma^l_0(\o)\hat\t_0+\Sigma_3^l(\o)\hat\t_3$ is 
the self-energy due to the OSSF defined in the main text. From now on we do not 
display the ($\bk,\o$) dependence.
We can define the renormalize Hamiltonian as $\hat H^l= \hat H^l_0 + \hat 
\Sigma^l $. Rotating the Hamiltonian $\hat H^l $  into the band basis we have 
$\hat H^l ={\red{\hat{\cal U}_R}^{l^{-1}}} 
\red{\hat{\Lambda}}_R^{l}{\red{\hat{\cal U}}_R}^l$ where 
$\red{\hat{\Lambda}}_R^{l} = diag (E_{R}^{l_+} , E_{R}^{l_-} )$ is the 
eigenvalue matrix and $E_{R}^{l_\pm} $ is given by 
\bea
E_{R}^{l_\pm}&=&h_0^l + \Sigma^l_0 \pm h_R^l=h_0^l + \Sigma^l_0 + \nn \\
&& \pm \sqrt{(h^l_{1})^2+(h^l_{2})^2 + (h^l_{3} + \Sigma_3^l)^2}.
\lb{h}
\eea
Since the self-energy given in Eq.~\pref{eq:Dyson} is complex, we define the 
renormalized energy dispersion relation as $\epsilon_{R}^{l_\pm} = Re 
E_{R}^{l_\pm}$ and the renormalized scattering rate as $\G_R^{l_\pm} = 
\cyan{\delta\G} + |Im E_{R}^{l_\pm}|$ \red{with $\delta\G $ some scattering 
background}. The unitary $\hat {\cal U}_R^l $ matrix has a structure analogous 
to Eq.~\pref{dd} provided that the coherence factors are given by 
\bea
\lb{i}
u^l_R &=&\frac{1}{\sqrt{2}} \sqrt{1+\frac{h_{3}^l + \Sigma_3^l}{h_R^l}}, \nn \\  \nn
v^{*l}_R &=&\frac{1}{\sqrt{2}}\frac{h_{1}^l + ih_{2}^l}{\sqrt{(h_{1}^l)^2 + 
(h_{2}^l)^2}} \sqrt{1-\frac{h_{3}^l + \Sigma_3^l}{h_R^l}}.\\
\eea
We can take into account in our model the presence of a finite spin-orbit 
interaction by replacing $h_R^\G \to h_{R, SO}^\G=\sqrt{(h^\G_{1})^2 + (h^\G_{3} 
+ \Sigma_3^\G)^2+(\lambda^2/4)}$. This affects both the $E_R^{\G_\pm}$ and the 
coherence factors $(u/v)^{\G}_R$.
\subsection*{Perfectly nested circular Fermi surfaces model}
For the analytical approach, we write the bare pocket Hamiltonian Eq.~\pref{a} 
in polar coordinates with $\theta=\arctan k_y/k_x$, for the hole 
\bea
h_0^{\G} &=& \epsilon^{\G} -ak^2; \nn \\ 
h_1^{\G} &=& -bk^2\sin(2\theta); \nn \\ 
h_3^{\G} &=& bk^2\cos(2\theta) 
\lb{j1}
\eea
and for the electron pockets
\bea
h_0^{X/Y} &=& -\epsilon^{X/Y} +ak^2; \nn \\
h_2^{X/Y} &=& bk^2\sin(2\theta); \nn \\ 
h_3^{X/Y} &=& \mp bk^2\cos(2\theta) 
\lb{j2}
\eea
For simplicity let us further assume $\epsilon^X \equiv \epsilon^Y \equiv 
\epsilon^e$. It is easy to check that the orbital content of the 4-pocket model 
for IBS in Fig.~\ref{fig-OSSF} is reproduced in the parabolic approximation. 
Using Eq.~\pref{c} the coherence factors take the expression 
\bea
\lb{nn}
\begin{split}
|u^\G|^2&=|u^Y|^2=|v^X|^2=\cos \theta^2,  \\ 
|v^\G|^2&=|v^Y|^2=|u^X|^2=\sin \theta^2  
\end{split}
\eea
To allow analytical treatment, we expand $E_{R}^{l_\pm}$ and $(u/v)^{l}_R$ 
Eq.\pref{i} Eq.\pref{h} up to first order in $\Sigma_3$ and we neglect the 
imaginary part of the self-energy since goes to zero at low $\omega$ and T = 0. 
The dressed coherence factors are given by: 
\bea
\lb{coherencef}
\begin{split}
|u^l_{R}|^2=&|u^l|^2 \left(1 + \frac{2{Re}\Sigma_3^l}{h^l}|v^l|^2\right), \\
|v^l_{R}|^2=&|v^l|^2 \left(1 - \frac{2{Re}\Sigma_3^l}{h^l}|u^l|^2\right)
\end{split}
\eea
that are the expressions quoted in the main text in Eq.~\pref{uvrenexpand}. The 
dressed dispersion relations $\epsilon_{R}^{l_\pm} = Re E_{R}^{l_\pm}$ become:
\be
\lb{k}
\epsilon_{R}^{l_\pm} = \epsilon^{l_\pm}  + Re\Sigma^l_0 \pm \frac{h^l_{3}}{h^l} Re\Sigma_3^l.
\ee
Replacing the values for the case of circular FS given in 
Eq.~\pref{j1} and Eq.~\pref{j2}  and calculate it at the Fermi surface we get: 
\bea
\lb{l}
&\epsilon_{R}^{\G_\pm} &= \ \epsilon_0^{h}  \pm \cos 2\theta Re\Sigma_3^\G, \nn \\ 
&\epsilon_{R}^{{X/Y}} &= \ \red{-}\epsilon_0^{e} \mp \cos 2\theta Re\Sigma_3^{X/Y}, 
\eea
where we defined the tetragonal band energy for the h-pockets as $\epsilon_0^{h} 
= \epsilon^\G + Re\Sigma^\G_0$ and $\epsilon_0^{e} = \epsilon^e \red{-} 
Re\Sigma^{X/Y}_0$ for the e-pockets. 
The scattering rate $\G_R^{l_\pm} = \delta\G + |Im E_{R}^{l_\pm}|$  acquires the 
expression: 
\be
\lb{q}
\G_{R}^{l_\pm} \approx \delta\G + |Im\Sigma^l_0| \pm \frac{h^l_{3}}{h^l} 
|Im\Sigma_3^l|,
\ee 
Using Eq.~\pref{j1} and Eq.~\pref{j2} we get:
\bea
\lb{ll}
&\G^{\G_\pm}_R& =\  \G^h_0 \pm \cos 2\theta |{Im}\red{\Sigma_3}^\G|, \nn \\ 
&\G^{X/Y}_R& =\  \G^{e}_0 \mp \cos 2\theta |{Im}\red{\Sigma_3}^{X/Y}|,
\eea
where we have separated the angular dependent renormalization $\sim 
Im\Sigma^l_3$ from the tetragonal constant part, $\G^{h/e}_0=\delta\G 
+|{Im}\Sigma_0^{\G/e}|$. From Eq.~\pref{ll} we find analytically the 
locations of the cold spots on the FS where the minimum value of the scattering 
rate is found:
\bea
\G^{\G_+}_R(\theta=\frac{\pi}{2})&=&   \delta\G+ |{Im}\Sigma^\G_{xz}|,  \nn \\
\G^{\G_-}_R(\theta=0)&=&  \delta\G + |{Im}\Sigma^\G_{xz}| , \nn \\
\G^{X/Y}_R (\theta=0)&=& \G^{Y}_R(\theta= \frac{\pi}{2}) = \delta\G  
\eea 
This result is easy to understand due to the OSSF shown in Fig.~\ref{fig-OSSF}
where cold spots are also shown. In the nematic phase, SF with momentum
$\red{\bold{Q}}_X$ are bigger than the ones with $\red{\bold{Q}}_Y$. As a
consequence, the largest scattering is found in the $yz$ orbital due to the
orbital selective nature of SF, while the $xy$-orbital component is not
renormalized given the absence of $xy$-OSSF.

\section*{\bf{APPENDIX B: Computation of the DC conductivity and analytical 
results for perfectly nested circular FS}} 
We calculate the DC conductivity in 
the bubble approximation. The self-energy
corrections computed within an Eliashberg-like treatment are momentum independent so
that vertex corrections vanish identically. The DC conductivity is given by
\be
\lb{aa}
\text{Re}\,\sigma_{\a}(\O)=-\frac{e^2}{V}\frac{\text{Im}\, 
\Pi_{\a}(\bold{q}=0, i\O_m \ra \O)}{\O}
\ee
where $\a = x,y$, V is the unit-cell volume and $\Pi_\a$ is the current-current 
correlation function in the bubble approximation given by
\be
\lb{bb}
\Pi_\a({i\O_m})=2T \sum_{l \bk n} \text{Tr}\left[\hat{G\red{^l}}(\bk,i\o_n)
\cyan{ \hat {V}_{\bk_\a}^l} \hat G\red{^l} (\bk,i\o_n+i\O_m) \hat {V}_{\bk_\a}^l \right].
\ee
$\hat {V}_{\bk_\a}^l=\partial_{\bk\red{_\a}} \hat H_0^l$ is the bare 
velocity operator and $\hat{G\red{^l}}(\bk,i\o_n)$ is the renormalized Green's 
function from the orbital-selective spin fluctuations. We work in the rotated 
basis and replace $\red{\hat G^{l}}(\bk,i\o_n)={\red{\hat{\cal 
U}_R^l}}(\bk,i\o_n)(i\o_n\red{\hat{\mathbb{1}}}-\red{\hat{\Lambda}}_{R}^l(\bk,
\o))^{-1}\red{{\hat{\cal U}}_R^{l^{-1}}}(\bk,i\o_n)$ in Eq.~\pref{bb}. 
Since we are interested in the DC conductivity ($\O \ra 0$) the important term 
in the trace is 
\bea
\lb{trace}
\text{Tr}&\big[&{\hat{\cal U}_R^l}(\bk,i\o_n)
(i\o_n\hat{\mathbb{1}}- \hat{\Lambda}_{R}^l(\bk,\o))^{-1} 
{\hat{\cal U}_R^{l^{-1}}}(\bk,i\o_n) \hat {V}_{\bk_\a}^l  \nn \\
&& {\hat{\cal U}_R^l}(\bk,i\o_n)(i\o_n\hat{\mathbb{1}}-  
\hat{\Lambda}_{R}^l(\bk,\o))^{-1}{\hat{\cal U}}_R^{l^{-1}}(\bk,i\o_n)
\hat{V}_{\bk_\a}^l \big] \nn 
\eea
Using the cyclic property of the trace allows us to define the renormalized 
velocity $\red{\hat V}_{R\bk_\a }^{l}=\red{{\hat{\cal U}}_R^{l^{-1}}}(\bk,i\o_n) 
\cyan{ \hat {V}_{\bk_\a}^l}{\red{\hat{\cal U}_R^l}}(\bk,i\o_n)$. Operating the 
trace and taking the limit $\O \to 0$ we get that the trace in Eq.~\pref{trace} 
can be written as
\bea
\begin{split}
g_{+}^2 (\bk,i \o_n) \red{(V_{R\bk_\a}^{l_+})^2} \ +
\ g_{-}^2 (i\o_n) (V_{R\bk_\a}^{l_-})^2
\end{split}
\lb{a.45}
\eea
where $V_{R\bk_\a }^{l_\pm} $ are given by 
\be
\lb{ddd}
\red{V}^{l_\pm}_{R\bk_\a }= \cyan{V_{\bold{k}\a}^{l_{11}}} |u^l_{R}|^2 
\pm \cyan{V_{\bold{k}\a}^{l_{12}}} u^{*l}_{R}v^{*l}_{R} \pm 
V_{\bold{k}\a}^{l_{21}} u^l_{R}v^l_{R} +\cyan{V_{\bold{k}\a}^{l_{22}}} |v^l_{R}|^2
\ee
Here $V_{\bold{k}\a}^{l_{\eta \eta^{\prime}}}$ are the $\eta \eta^{\prime}$
component of $\hat {V}_{\bk_\a}^l=\partial_{\bk\red{_\a}} \hat H_0^l$ and
$({u/v})^l_{R}$ the renormalized  coherence factors defined in Eq.~\pref{i}. The
multiorbital character of the problem gives rise to self-energy effects in the
velocities via the coherence factors especially in the nematic phase. Expressing
the Green's function in terms of the spectral functions $A^{l_\pm}(\bk, \o)$ we
finally arrive to the results of Eq.~\pref{eq:DC} quoted in  main text: 
\be
\lb{sigma}
\cyan{\sigma^{l_\pm}_{\a}} = \frac{2 \pi e^2}{ N}\sum_{\bk}\int_{-\infty}^{\infty}
d\omega(-\frac{\partial f(\omega)}{\partial\omega}) 
\big({V}_{R\bk_\a }^{l_\pm}(\o)\big)^2 \big( A^{l_\pm}_\bk(\o)\big)^2,
\ee
where $f(\omega)$ the Fermi distribution function. Notice that the renormalized 
velocities $V_{R\bk_\a }^{l_\pm}(\o)$ depend on frequency and on the orbital 
self-energy through the dependence of the coherence factors on $\Sigma^l_3(\o)$. 
The renormalized spectral function can be written as 
\begin{equation}
\lb{ff}
A^{l_\pm}_\bk(\o)=\frac{1}{\pi}\frac{\G_{R\bk}^{l_\pm}(\o)}{(\G_{R\bk}^{l_\pm}(\o))^2 + 
(\omega - \epsilon_{R\bk}^{l_\pm}(\o))^2}
\end{equation}
where $\epsilon_{R\bk}^{l_\pm}(\o)  = Re E_{R}^{l_\pm}(\bk,\o)$ and  
$\G_{R\bk}^{l_\pm}(\o) = \delta\G +|Im E_{R}^{l_\pm}(\bk,\o)|$. At low temperature 
$ -\frac{\partial f(\omega)}{\partial\omega} \rightarrow 
 \delta(\omega)$ that selects only states at the Fermi level 
\be
\lb{hhh}
\sigma^{l_\pm}_{\a}=\frac{2\pi e^2}{N}\sum_{\bk} ({V}_{R\bk_\a }^{l_\pm})^2 
\big( A^{l_\pm}_\bk\big)^2
\ee
Moreover, assuming $\G_R^{l_\pm} \rightarrow 0 $, the spectral function 
$A^{l_\pm}_\bk$ can be approximated as 
\begin{equation}
\lb{AproxAR}
\big( A^{l_\pm}_\bk\big)^2 \longrightarrow \frac{1}{2 \pi \G_{R\bk}^{l_\pm}}  \ 
\delta (\epsilon_{R\bk}^{l_\pm})
\end{equation}
Replacing this expression in Eq.~\pref{hhh} we get that the conductivity is
\bea
\lb{DC2}
\sigma^{l_\pm}_{\a}&=&\frac{e^2}{N}\sum_{\bk} 
\frac{(\red{V}_{R\bk_\a }^{l_\pm})^2}{\G_{R \bk}^{l_\pm}}  \ 
\delta (\epsilon_{R \bold{k}}^{l_\pm}) =\\ \nonumber 
&=& \frac{e^2}{N}\int \frac{d \bk^2}{(2\pi)^2} 
\frac{(V_{R\bk_\a }^{l_\pm})^2}{\G_{R \bk}^{l_\pm}}  \ 
\frac{\delta (\bk -\bk_F)}{| \bigtriangledown \epsilon_{R \bk}^{l_\pm}|}
\eea
%

Within the parabolic band approximation and using the expansion up to the first
order in $\Sigma^l_3$ of the renormalized energy, we can derive analytically the
conductivity given in Eq.~\pref{DC2}. We use the expressions of the renormalized
energies and scattering rates derived in App.\,A. Using the expressions
Eq.~\pref{coherencef} for the coherence factors, it is easy to check the
velocity given in Eq.\pref{ddd} can be expressed as the derivative of the
renormalized dispersion relation given in Eq.~\pref{k}, so $V^{l\pm}_{R \bk
\a}=\partial_{k_\a} \epsilon_R^{l_{\pm}} (\bk)$. Explicitly for the h-pockets at
$\Gamma$ we have 
\bea
\lb{hh}
V^{\G_\pm}_{R \bk \a} &=&  \partial_{k_\alpha} h_0^\G \pm\partial_{k_\alpha} \, 
h_1^\G \frac{h_1^\G}{h^\G}\pm \partial_{k_\alpha} \, h_3^\G \frac{h_3^\G}{h^\G}  \nn \\
&\pm& {Re}\Sigma^\G_3 \frac{h_1^\G}{(h^\G)^2} 
\Big[\partial_{k_\alpha}h_3^\G\frac{h_1^\G}{h^\G}-\partial_{k_\alpha} 
h^\G_1 \frac{h_3^\G}{h^\G}  \Big]
\eea
and analogous expressions for $V^X_{R \bk \a}$ and $V^Y_{R \bk \a}$. The first
three terms in Eq.~\pref{hh} corresponds to the bare velocity, while the term
multiplied by ${Re}\Sigma^\G_3 $ accounts for the renormalization in the
velocity due to OSSF self-energy corrections. Using the explicit definition of
$h^l_0$ and $\vec{h}^l$, the velocities for the various pockets read: 
\bea
\lb{velorenortodospockets}
\red{V}^{\G_\pm}_{Rx}&=&\cyan{ -\frac{k \cos\theta }{m^{\G_\pm}} 
\pm 4 Re\Sigma^{\G}_3\frac{k \cos\theta\sin\theta^{2}}{k^{2}} },\nonumber  \\ 
\red{V}^{\G_\pm}_{Ry}&=&\cyan{  -\frac{k \sin\theta }{m^{\G_\pm}} 
\mp 4 Re\Sigma^{\G}_3\frac{k \sin\theta\cos\theta^{2}}{k^{2}} },\nonumber  \\ 
\red{V}^{X/Y}_{Rx}&=&\cyan{ \frac{k \cos\theta }{m^{X/Y}} 
\mp 4 Re\Sigma^{X/Y}_3\frac{k \cos\theta\sin\theta^{2}}{k^{2}} },  \nonumber\\ 
\red{V}^{X/Y}_{Ry}&=&\cyan{ \frac{k \sin\theta }{m^{X/Y}} 
\pm 4 Re\Sigma^{X/Y}_3\frac{k \sin\theta\cos\theta^{2}}{k^{2}} },
\eea
where Eq.~\pref{j1} and Eq.~\pref{j2} have been used and $m^{l_\pm}$ is the bare
mass of the $l_\pm$ pocket whose definition in terms of the Hamiltonian
parameters is given by $m^{\G_\pm} = {2(a \mp b)}^{-1}$ and $m^{X/Y} = m^{e}
={2(a+b)}^{-1}$.   
To compute the k integration in the Eq.~\pref{DC2} we evaluate the velocity in
the Eq.~\pref{ii} at the renormalized FS. In the nematic phase $k_{F}(\theta)$
is no longer constant. Using again the expansion up to the first order in
$\Sigma_3$ of the renormalized energy we can estimate $k_{F}(\theta)$ and
replace it in Eq.~\pref{ii}. We find
\bea
\lb{ii}
{V}^{\G_\pm}_{Rx}&=&\cyan{{V_0}^{\Gamma_\pm}_x 
\bigg( 1 \pm  \cos 2\theta \frac{Re\Sigma^{\G}_3}{2 \epsilon^{h}_0} 
\mp 4 \sin^2 \theta \frac{Re\Sigma^{\G}_3}{2 \epsilon^{h}_0}  \bigg) },\nonumber  \\ 
V^{\G_\pm}_{Ry}&=&\cyan{{V_0}^{\Gamma_\pm}_y 
\bigg( 1 \pm  \cos 2\theta \frac{Re\Sigma^{\G}_3}{2 \epsilon^{h}_0} 
\pm 4 \cos^2 \theta \frac{Re\Sigma^{\G}_3}{2 \epsilon^{h}_0}  \bigg) },\nonumber  \\ 
V^{X/Y}_{Rx}&=&\cyan{{V_0}^{X/Y}_x 
\bigg( 1 \pm  \cos 2\theta \frac{Re\Sigma^{X/Y}_3}{2 \epsilon^{X/Y}_0} 
\mp 4 \sin^2 \theta \frac{Re\Sigma^{X/Y}_3}{2 \epsilon^{X/Y}_0}  \bigg) },  \nonumber\\ 
V^{X/Y}_{Ry}&=&\cyan{{V_0}^{X/Y}_y 
\bigg( 1 \pm  \cos 2\theta \frac{Re\Sigma^{X/Y}_3}{2 \epsilon^{X/Y}_0} 
\pm 4 \cos^2 \theta \frac{Re\Sigma^{X/Y}_3}{2 \epsilon^{X/Y}_0}  \bigg) }.\nonumber\\
\eea
where  $\epsilon^{h}_0$ and $\epsilon^{e}_0$ are the Fermi energy in the
tetragonal phase defined in Eq.s~\pref{l} and  \red{${V_0}^{\G_{\pm}}_\a= -
{k_0}_{F \a}^{\G_{\pm}}/ m^{\G_{\pm}}$, ${V_0}^{X/Y}_\a= {k_0}_{F \a}^{e}/
m^{e}$ } is the $\a$ component of the bare velocity with \red{
${k_0}_{F}^{\G_{\pm}} = \sqrt{ \epsilon^{h}_{0}/(2m^{\G_{\pm}})}$ and $
{k_0}_{F}^{e} = \sqrt{ \epsilon^{e}_{0}/(2m^{e}})$}. 

The last term we need to evaluate is the $| \bigtriangledown 
\epsilon_{R\bk}^{l_\pm}|$ (see Eq.~\pref{DC2}). In the $\G$ pocket turns out to be 
$|\bigtriangledown \epsilon_{R\bk}^{l_\pm}|=k / m^{\G_{+}}$ with similar results 
for the other pockets. The important point is that the norm of the pocket 
velocity is independent of the self-energy while the pocket velocity in a given 
direction $x$/$y$ depends on the self-energy, Eq.~\pref{ii}. 
Replacing all the analytical expressions found for the velocities, Eq.~\pref{ii}, 
and the scattering rate, Eq.~\pref{ll} in Eq.~\pref{DC2} the pockets DC 
conductivities read: 
\bea
\s_{x/y}^{\G_+}&=&\s^{h}\bigg(1\pm \frac{{Re}\Sigma_3^{\G}}{2\epsilon^h_0} 
\mp \frac{{Re} \Sigma_3^{\G}}{\epsilon^h_0} \mp \frac{|{Im}\Sigma^\G_3|}{2\G^{h}_0}\bigg) \nn \\
\s_{x/y}^{\G_-}&=&\s^{h}\bigg(1 \mp \frac{{Re}\Sigma_3^{\G}}{2\epsilon^h_0} 
\pm \frac{{Re} \Sigma_3^{\G}}{\epsilon^h_0} \pm\frac{|{Im}\Sigma^\G_3|}{2\G^{h}_0}\bigg) \nn \\
\s_{x/y}^{X}&=&\s^e\bigg(1\pm \frac{{Re}\Sigma_3^{X}}{2\epsilon^e_0} 
\mp \frac{{Re} \Sigma_3^{X}}{\epsilon^e_0} \pm \frac{|{Im}\Sigma^X_3|}{2\G^{e}_0}\bigg) \nn \\
\s_{x/y}^{\red{Y}}&=&\s^e\bigg(1\mp \frac{{Re}\Sigma_3^{Y}}{2\epsilon^e_0} 
\pm \frac{{Re} \Sigma_3^{Y}}{\epsilon^e_0} \mp \frac{|{Im}\Sigma^Y_3|}{2\G^{e}_0}\bigg)
\eea
where $\sigma^{h/e}= e^2\epsilon^{h/e}_0/(2 \pi \hbar)\G_0^{h/e}$ is the 
$x/y$-component of the DC conductivity in the tetragonal phase for the 
h/e-pocket. 
The DC conductivity anisotropy for the h- and e-pockets can be finally written 
as
\bea
\Delta \sigma_{DC}^{h+} &=& \s^{\G} \left(\frac{\Phi^h}{\epsilon^h_0} 
-\frac{\Delta \G^h}{\G^h_0} \right) 
\nn \\
\Delta \sigma_{DC}^{h-} &=& \s^{\G} \left(-\frac{\Phi^h}{ \epsilon^h_0} 
+\frac{\Delta \G^h}{\G^h_0} \right)  
\nn \\
\Delta \sigma_{DC}^{e} &=&  \s^{\red{e}} \left(- \frac{\Phi^e}{\epsilon^{\red{e}}_0}  
+ \frac{\Delta \G^e}{\G^{\red{e}}_0} \right) 
\eea
with the nematic order parameters for the h-pockets ($\Phi^h, 
\Delta\Gamma^h$) and the e-pockets ($\Phi^e, \Delta\Gamma^e$) defined as 
\bea
\Phi{^h} \equiv \displaystyle\frac{Re\Sigma^{\G}_{xz}-Re\Sigma^{\G}_{yz}}{2}=
-Re \Sigma^{\G}_3 ,
\quad  \Delta\Gamma{^h}\equiv|Im \Sigma^{\G}_3|,  \nn \\
\Phi{^e}\equiv \frac{Re\Sigma^X_{yz}-Re\Sigma^Y_{xz}}{2} ,
\quad   \Delta\Gamma{^e}\equiv \frac{|Im \Sigma^X_{yz}|
-|Im \Sigma^Y_{xz}|}{2} \nn  \\
\eea

\section*{APPENDIX C: Model parameters for FeSe and \\ 122 systems}
To perform the numerical analysis discussed in the main text we used a set of 
band parameters which reproduce the experimental dispersions observed in 122 and 
FeSe systems. Using the band parameters listed in Table \ref{bp}, a spin-orbit 
interaction of $\lambda= 5$ meV and $|Re\Sigma_{yz/xz}^\G|= Re\Sigma_{yz/xz}^{X/Y}= 
15$ meV, we obtain the FS topology shown in Fig.1 that reproduce qualitatively 
well the FS of 122 systems in the tetragonal phase. The nematic phase is 
computed with a symmetric nematic splitting of $\Phi^h = \Phi^e= 4$ meV. 
For the FeSe case we used the same set of band parameters listed in Table 
\ref{bp}, with spin-orbit interaction $\lambda= 20$ meV, and 
$|Re\Sigma_{yz/xz}^\G| = 70/40$ meV and $Re\Sigma_{yz/xz}^{X/Y} = 45/15$ meV that 
result in a nematic order parameter $\Phi^{h/e} = 15$ meV. This set reproduce 
the FS and their orbital distribution as experimentally observed by 
ARPES\cite{Fanfarilloprb16, ColdeaWatsonreview18, Rhodesprb18} in the nematic 
phase. 

\begin{table}[tbh]
\begin{center}
\begin{tabular}{ccccccccccccccccccc}
\hline 
&              &    $\Gamma$   &   &\qquad \qquad \qquad&                       &X&                     &\\
\hline \hline
&$\epsilon_\G$ &               & 46&                    &$\epsilon_{xy}$ \ \ 72 & &$\epsilon_{yz}$\ \ 55&\\
\hline
& $a_\G$       &               &263&                    & $a_{xy}$\ \ 93        & &$a_{yz}$\ \ 101       &\\
\hline
& $b_\G$       &               &182&                     & $b$ \ \ \ 154          & &                     &\\
\hline
&              &               &   &                    & $v$ \ \ \ 144          & &                     & \\
\hline
\hline
\end{tabular}
\caption{Low-energy model parameters used for FeSe and 122 system. 
All the parameters are in meV, the $k$ vector is measured in units $1/a \sim 
0.375$ \AA, where $a=a_{FeFe}$ is the lattice constant of the 1-Fe unit cell (so 
that $\tilde a=\sqrt{2}a=3.77$ \AA \, is the lattice constant of the 2Fe unit 
cell).} 
\vspace{-0.5cm}
\label{bp}
\end{center}
\end{table} 

We fix the background scattering to $\d \G = 1$ meV. The scattering rates used 
in the tetragonal phase for 122 are $Im\Sigma^\G_{yz/xz}= Im 
\Sigma_{yz/xz}^{X/Y} = -2$ meV, while for the FeSe case we used 
$Im\Sigma^\G_{yz/xz}= -4$ meV and $Im \Sigma_{yz/xz}^{X/Y} = -3$ meV.
In both cases their variations of the imaginary part of the self-energies in the 
nematic phase are assumed to be proportional to the variation of their real 
parts, i.e. $\D\G^h\sim c_h \ (\Phi^h/Re\Sigma_0^\G)$ and $\D\G^e\sim c_e \ 
(\Phi^e/Re\Sigma_0^{X/Y})$ with $c_{h/e}$ arbitrary coefficients. 

\bibliography{pnictides_nematic}

\end{document}